\newcommand\ignore[1]{}
\documentclass[12pt,citesort]{article}
\usepackage{amsmath,amssymb,graphicx,url}
\usepackage{color}
\newcommand\Black{}

\newcommand{\half}{ {\scriptstyle \frac{1}{2} } }
\newcommand{\Half}{ {\frac{1}{2} } }

\newcommand\be{\begin{equation}}
\newcommand\ee{\end{equation}}
\newcommand\bea{\begin{eqnarray}}
\newcommand\eea{\end{eqnarray}}
\newcommand{\bdm}{\begin{displaymath}}
\newcommand{\edm}{\end{displaymath}}
\newcommand\nn{ \nonumber\\}

\newcommand{\dd}[1]{\partial_{#1}}
\def\tr{{\rm Tr}}\def\dd{\partial}
\renewcommand{\>}{\rangle}

\makeatletter
\@addtoreset{equation}{section}
\makeatother
\renewcommand{\theequation}{\thesection.\arabic{equation}}
\title{Odderon in Gauge/String Duality}

\author{Richard  C. Brower\footnote{Physics Department,
Boston University, Boston MA 02215},
Marko Djuri\'c\footnote{Physics Department, Brown University,
Providence, RI 02912},  and  Chung-I Tan\footnotemark[\value{footnote}]
}

\begin{document}

\maketitle

\begin{abstract}
At high energies, elastic hadronic cross sections, such as $pp, \; \overline p   p, \; \pi^{\pm} p$, are dominated by vacuum exchange, which in leading order of   the $1/N_c$ expansion has been identified as the BFKL Pomeron or its strong   AdS dual the closed string Reggeized graviton~\cite{Brower:2006ea}.  However the   difference of particle anti-particle cross sections are given by a so-called   Odderon, carrying C = -1 vacuum quantum numbers identified in weak coupling   with odd numbers of exchanged gluons. Here we show that in the dual   description the Odderon is the Reggeized Kalb-Ramond field   ($B_{\mu\nu}$) in the Neveu-Schwartz sector of closed string theory.  To   first order in strong coupling, the high energy contribution of Odderon is   evaluated for ${\cal N} = 4$ Super Yang-Mills by a generalization of the gravity   dual analysis for Pomeron in  Ref.~\cite{Brower:2006ea}.  The consequence of confinement on the Odderon are   estimated in the confining QCD-like $AdS^5$ hardwall model of Polchinski and   Strassler~\cite{Polchinski:2001tt}.
\end{abstract}

\begin{flushright}
Brown-HET-1569
\end{flushright}
\vspace{-5cm}

\newpage

\setcounter{tocdepth}{3}
\tableofcontents

\newpage

\section{Introduction}

\begin{figure}[bthp]
\begin{center}
\includegraphics[width = .7\textwidth]{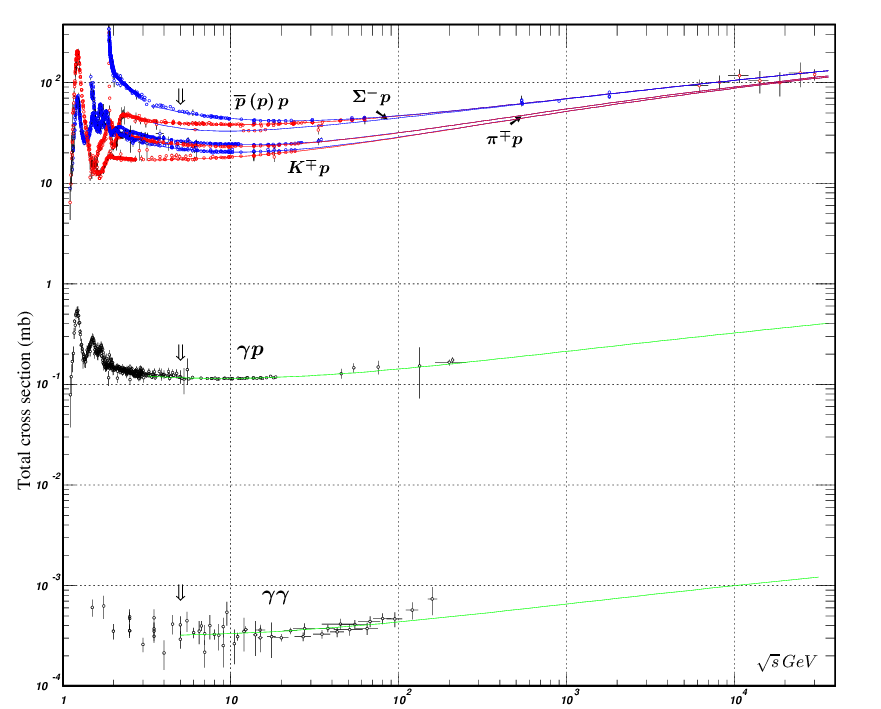}
\end{center}
\caption{The continued growth of total cross sections for $pp$, $\bar pp$,$\pi^{\pm} p$, etc.,  over a wide range of energies. }
\label{fig:crossX}
\end{figure}

\begin{figure}[bthp]
\begin{center}
\includegraphics[width = 1.0\textwidth]{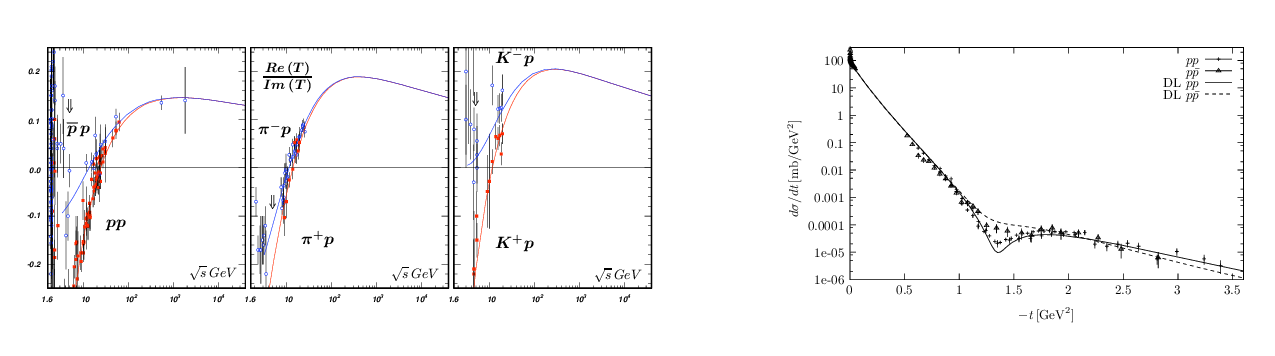}
\end{center}
\caption{Energy dependence of $\rho (s) = {\rm Re} F/{\rm}  Im F$ at $t\simeq 0$ can be used to infer the presence  of Odderon contributions \cite{Gauron:1986nk}.  The difference between differential cross sections for elastic $pp$ and $p\bar p$ at $\sqrt s = 53$ GeV in the ``dip region'' \cite{Donnachie:1983hf} has been advocated as the best evidence for the existence of Odderon. For a careful review, see \cite{Ewerz:2005rg}.}
\label{fig:rho-dip}
\end{figure}

One of the most striking aspects of high energy hadron-hadron scattering is the continued growth in the total cross section $\sigma_T$ from collider to cosmic ray energies. (See Fig.~\ref{fig:crossX}.) This increase can be fitted by a power $\sigma_T \sim s^{j^{+}_0 -1}$ with intercept, $j^{+}_0 >1$, to represent the so-called Pomeron Regge exchange in leading order in the $1/N_c$ expansion. Alternatively it can be fitted by $\sigma_T \sim log^2 s$, the maximally allowed asymptotic term consistent with saturating the Froissart unitarity bound. In either case, one also observes a significant component in the difference of the antiparticle-particle and particle-particle total cross sections.  This charge conjugation odd exchange, $C=-1$, which is referred to as the Odderon contribution \cite{Lukaszuk:1973nt,Bialkowski:1974cp,Finkelstein:1989mf,Avila:2006wy,Nicolescu:2007ji}, is often fitted by another sub-leading power $\Delta \sigma_T(s) \sim s^{j^{-}_0-1}$. The splitting between the two powers, $j^{+}_0 - j_0^{-} > 0 $, can  be inferred by the ratio of real/imaginary amplitudes as well as by differential cross sections in the near-forward limit.  (See Fig.~\ref{fig:rho-dip}.)  
We study the $C=-1$ $J$-plane singularities in the {\em crossing-odd} sector,  from the perspective of {\em Gauge/String Duality},  giving the first strong coupling evaluation of the Odderon in ${\cal N} = 4$ super Yang Mills theory
in the large $N_c$ 't Hooft limit. We find that, while the Pomeron emerges as fluctuations of the metric tensor, $G_{MN}$, the Odderon is  associated with fluctuations in  anti-symmetric tensor field, $B_{MN}$, the {\em Kalb-Ramond} (KR) fields \cite{Kalb:1974yc} in $AdS_5$ background.

Although the notion of a Pomeron has been around since the early sixties, its theoretical underpinning in a non-perturbative setting was   only  understood recently.  Brower, Polchinski, Strassler and Tan~\cite{Brower:2006ea} have shown that, for a conformal theory in the large $N_c$ limit, a dual Pomeron can always be identified with the leading eigenvalue of a Lorentz boost generator $M_{+-}$ \cite{Brower:2007xg}.  A related weak-strong extrapolation for ${\cal   N}=4$ Super YM has also been carried out in \cite{klv5}.  In the strong coupling limit, conformal symmetry~\footnote{ For the weak coupling BFKL, this   is referred as M\"obius invariance which in strong coupling is   realized~\cite{Brower:2007xg,Brower:2007qh} as the $SL(2,C)$ isometries of   Euclidean $AdS_3$ subspace of $AdS_5$. See also   \cite{Cornalba:2007fs,Cornalba:2006xm}.}  requires that the leading $C=+1$ Regge singularity is a fixed $J$-plane cut, which for ${\cal N} = 4$ super Yang Mills theory is located at
\be
j^{(+)}_0 = 2 - 2/\sqrt{\lambda}+O(1/\lambda)\; . \label{eq:dualconformalPomeron}
\ee
As the 't Hooft coupling, $\lambda = g^2 N_c$, increases, the ``conformal Pomeron'' moves to $j=2$ from below. In the limit $\lambda \to \infty$, the conformal Pomeron corresponds to the $AdS_5$ graviton. We extend in this paper the $C=+1$ analysis to the $C=-1$ sector. We demonstrate that the strong coupling conformal Odderon, like the Pomeron, is again a fixed cut in the $J$-plane but with its intercept determined by the AdS mass squared, $m^2_{AdS}$, of  Kalb-Ramond field,
\be
j_0^{(-)}=1- m^2_{AdS}/2\sqrt{\lambda} + O(1/\lambda) \;. \label{eq:dualconformalOdderon}
\ee
Two possible solutions are found: one solution has $m^2_{AdS, (1)} = 16$ and a second possible solution has $m^2_{AdS, (2)}=0$, or $j_0^{(-)} = 1$ to this order. 

At weak coupling the classic study of Balitsky, Fadin, Kuraev and Lipatov (BFKL) ~\cite{Lipatov:1976zz,Kuraev:1977fs,BL,Lipatov:1985uk,KirschLipat} evaluated the Pomeron contribution to leading order in $g^2_{YM}$ and all orders  in $g^2_{YM} log(s/s_0)$. In the conformal limit, both the weak-coupling BFKL Pomeron and Odderon correspond to $J$-plane branch points. For instance, for the BFKL Pomeron, the cut is located at $ j^{(+)}_0 = 1 + ( \ln 2) \; \lambda/ {\pi^2} +O(\lambda^2)\; , $ where $\lambda = g^2 N_c$ is the 't Hooft coupling. Under the same leading log approximation investigations for the Odderon in the weak coupling were first carried out by Bartels, Kwiecinski and Praszalowicz (BKP) \cite{Kwiecinski:1980wb,Bartels:1980pe}.  Interestingly, two leading Odderon solutions have been identified. Both are branch cuts in the $J$-plane.  One has an intercept slightly below 1, at $j^{(-)}_{0,(1)} $, \cite{Janik:1998xj,Braun:1998fs}, and the second, denoted by $ j^{(-)}_{0,(2)} $, \cite{Bartels:1999yt}, has an intercept precisely at 1~\footnote{The fact   that the second solution has $j^{(-)}_{0,(2)}=1$ up to $O(\lambda^3)$ in the   conformal limit was communicated to us by Cyrille Marquest.}. These are summarized in Table~\ref{tab:intercepts}.

To illustrate the difference between the  Pomeron ($C=+1$) and the  Odderon ($C=-1$) sectors, consider the path ordered trace operator, $Tr P_\sigma \exp[i \oint d\sigma x'_\mu(\sigma) A_\mu(x)]$, which is the
source for the close string on the boundary of $AdS_5$. On the
string world sheet, charge conjugation is given by parity  ($\sigma \rightarrow -\sigma$), so that the  $C = \pm 1$ sectors
correspond to  symmetric/anti-symmetric (or real/imaginary) combinations,
 $$Tr P_\sigma [e^{ i \oint d\sigma x'_\mu(\sigma) A_\mu(x)} 
\pm e^{ - i \oint d\sigma x'_\mu(\sigma) A_\mu(x)}] \;, $$
respectively. Expanding to lowest order in perturbation theory, these reduce to two- and three-gluon source, where charge conjugation is $A_\mu(x) \rightarrow - A^T_\mu(x) $ with $A_\mu = \lambda^aA^a$, the Hermitian gauge field generators.  Thus the Pomeron Green's function for exchanging two gluons between hadrons can be written as a correlator, $\langle {\cal   P}_{\mu\nu}(x,y) {\cal P}_{\mu'\nu'}(x',y')\rangle$, of two color-singlet operators,
\be
{\cal P}_{\mu\nu}(x,y) = tr (A_\mu(x) A_\nu(y) ) = (1/2) \delta_{ab} A^a_\mu(x) A^b_\nu(y)
\ee
with $C=+1$.  For three gluons exchange, the Green's function can again be written as a correlator of color-singlet operators, $\langle {\cal O}_{\mu\nu\rho}(x,y,z) {\cal O}_{\mu'\nu'\rho'}(x',y',z')\rangle$. Unlike the case of two gluons, we now have two possibilities. One involves the totally symmetric $d$-coupling,
\bea
 {\cal O}_{\mu\nu\rho}(x,y,z) &=&   tr\left(\{  A_\mu(x), A_\nu(y)\}\; A_\rho(z)\right)= (1/2) d_{abc} A^a_\mu(x) A^b_\nu(y)A^c_\rho(z)
\eea
which is odd under C and  therefore is the lowest order contribution to the  Odderon.   The second possibility  involves the totally  anti-symmetric coupling,  $ \; tr\left([  A_\mu(x), A_\nu(y)] \; A_\rho(z)\right)= (i/2) f_{abc} A^a_\mu(x) A^b_\nu(y)A^c_\rho(z)\;. $  Under $C$, $f_{abc}\rightarrow f_{cba}=-f_{abc}$  and C even, leading  
to a three gluon contribution to the Pomeron.

In the weak coupling leading logarithmic approximation, the $C=+1$ BFKL Pomeron and the $C=-1$ Odderon~\cite{Kwiecinski:1980wb,Bartels:1980pe} are given by the ladder approximation  for two and three Reggeized gluons in  the t-channel.  Rapid advances have been made recently by exploring the consequences of conformal invariance. It is possible to treat each sector with fixed number, $n_g$, of Reggeized gluons exchanges in the large $N_c$ limit, $n_g=2,3,\cdots$. The leading intercept for each $n_g$-sector can be found as the ground state eigenvalue of a Hamiltonian, $H_{\rm BFKL}$, where
\be
H_{\rm BFKL}
= \frac{1}{4}\sum^{n_g}_{i=1}[{\cal H}(J^2_{i,i+1}) + {\cal H}(\bar
J^2_{i,i+1})],
\ee
is  a sum over two-body operator with holomorphic and
anti-holomorphic functions of the Casimir.  It has been shown in \cite{Lipatov:1994xy}  that this is a spin chain with 
a sufficient number of conserved charges to be completely integrable. In \cite{Faddeev:1994zg}  it 
was then  identified as an $SL(2,C)$ XXX Heisenberg spin chain. For the $n_g = 3$ Odderon, explicit solutions were found in \cite{Janik:1998xj,Braun:1998fs,Bartels:1999yt}. (For a comprehensive review on Odderon, see \cite{Ewerz:2005rg}. Other related works  include
\cite{Hatta:2005as,Kovchegov:2003dm,Kovner:2005qj}.)  
\begin{table}[h]
\begin{center}
\begin{tabular*}{150mm}{@{\extracolsep\fill}||c||l|ll||}
\hline
\hline
 & \multicolumn{1}{c|}{Weak Coupling}   &
 \multicolumn{1}{c}{Strong Coupling}          &    \\
\hline\hline\hline
$\;\;\; C=+1\;\;\;$ & $j^{(+)}_0 = 1
+  ( \ln 2) \; \lambda/ {\pi^2}  +O(\lambda^2)$    
   & $j^{(+)}_0 = 2 - 2/\sqrt{\lambda}+O(1/\lambda)$            &        \\
\hline\hline
$C=-1$    &   $j^{(-)}_{0,(1)} \simeq  1- 0.24717\;  \lambda/\pi + O(\lambda^2)$                                      
     & $j^{(-)}_{0,(1)}=1- 8/\sqrt{\lambda} + O(1/\lambda)$         &               \\
       &     $ j^{(-)}_{0,(2)} = 1 + O(\lambda^3)$              
      & $j^{(-)} _{0,(2)}=1+ O(1/\lambda)$                 &     \\
\hline
\hline
\end{tabular*}
\caption{Pomeron and Odderon intercepts at weak and strong coupling.}\label{tab:intercepts}
\end{center}
\end{table}

The AdS/CFT dictionary relates string states to local operators in the gauge theory. This can be summarized by the statement that gauge theory correlators can be found, in the strong coupling, by the behavior of bulk (gravity) fields, $\phi_i(x,z)$, as they approach the $AdS_5$ boundary, $z\rightarrow 0$,
\be
\langle e^{\int d^4x \phi_i(x) {\cal O}_i(x)}\rangle_{CFT} = {\cal Z}_{string} \left[ \phi_i(x,z) |_{z \sim 0}  \rightarrow   \phi_i(x) \right],
\ee
where $z>0$ extends into the $AdS$ bulk and the left hand side is the generating function of correlation functions in the gauge theory for the operator ${\cal O}_i$.  The Maldacena conjecture asserts that IIB superstring theory on $AdS_5\times S^5$ is dual to the ${\cal N}=4$ SYM conformal field theory on the boundary of the $AdS$ space.  For QCD, we need to adopt a metric which is asymptotically $AdS_5$ near the $AdS$ boundary (up to asymptotic freedom logarithmic corrections) and deformed in the infrared to model confinement.  Since  much  of the physics we will study takes place in the conformal region, for simplicity we begin by considering the $AdS_5\times S^5$ metric, obtaining first conformal results that strictly apply to ${\cal N} = 4$ super Yang Mills theory.   Furthermore,  as explained in \cite{Brower:1999nj,Brower:2000rp}, for the purpose of enumerating degrees of freedom in a gravity dual picture, it is sufficient to count modes at the $AdS$ boundary.  To address the effect of confinement, we consider only the hard-wall cut-off  introduced by Polchinski and Strassler~\cite{Polchinski:2001tt} as a toy model for   confining theories.

To be more precise, we consider IIB superstring theory, which, at low energy, has a supergravity multiplet with several bosonic fields: a metric tensor, $G_{MN}$, a dilaton $\phi$, an axion (or zero form R-R field) $C_0$, and the NS-NS and R-R fields $B_{MN}$ and $C_{2,MN}$ respectively.  To identify the relevant gauge-bulk couplings, we consider an effective Born-Infeld (BI) action plus Wess-Zumino term, describing the coupling of supergravity fields to a single D3-brane,
\be
S=\int d^4x
\det[G_{\mu\nu}+e^{-\phi/2}(B_{\mu\nu}+F_{\mu\nu})]+\int d^4x (C_0
F\wedge F+ C_2\wedge F+ C_4)\> \; ,
\ee
where $\mu, \nu\ = 1,2, 3,4$.     Both $B$ and $C_2$ are anti-symmetric tensors, and will be referred to as Kalb-Ramond fields. (In addition there is the 4-form RR field $C_{4}$ that is
constrained to have a self-dual field strength, $F_{5} = dC_{(4)}$.) 
 From the BI action, one finds that  the metric fluctuations couple to the energy-momentum tensor, which corresponds to   $C=+1$. For the $C=-1$ sector,  we find that  $B$ leads to  Odderon with even parity, $P=+ 1$ and $C_2$ leads to  Odderon with $P=-1$.

In Sec. \ref{sec:review}, we provide some general remarks on applying AdS/CFT to high energy scattering in QCD. This also serves to establish notations as well as to introduce some unavoidable background materials.  Readers familiar with these materials can move directly to the following sections. In Sec. \ref{sec:firstlook}, we discuss the Odderon in gauge/string duality from a target space perspective. We begin first with a qualitative discussion under an ``ultra-local'' approximation. Just like the case for $C=+1$, this also leads to a ``red-shifted'' Odderon. This effective Odderon is linear for $t>0$ and is a constant near $j=1$ for $t<0$, with a kink at $t=0$. We next introduce diffusion, moving from the flat-space scattering to AdS.  To simplify the discussion, we will focus on the conformal limit.  By turning the problem into an equivalent Schr\"odinger problem, the $J$-plane singularity follows from a standard spectral analysis. In particular, in the conformal limit, the spectrum consists a continuum, corresponding to a $J$-plane branch-cut, (\ref{eq:dualconformalOdderon}).  In Sec. \ref{sec:geo}, we provide a more formal interpretation of this finding from the perspective of $SL(2,C)$ invariance. In Sec. \ref{sec:Odderon}, we turn to a world-sheet analysis by providing a more systematic treatment of string scattering at high energies. Using OPE, we introduce vertex operators for both $C=\pm 1$, again moving from flat-space to AdS. We also provide a more general discussion on the connection between conformal dimensions and BFKL/DGLAP operators.  We discuss in Sec.  \ref{sec:spectrum} the effect of confinement deformation. We also enumerate $J^{PC}$ assignments for glueballs states, and point out the expected $J$-plane structure under confinement deformation.  We summarize and comment in Sec. \ref{sec:conclusion} on various related issues. We provide a short discussion on the second Odderon solution, $j^{(-)}_{0,(2)} =1$, from the strong coupling perspective, and comment on eikonalization for $C=\pm 1$ sectors.

\newpage

\section{Review of Gauge/String Duality in the Regge Limit}
\label{sec:review}

Before proceeding  to see how the curved-space analysis  for $C=+1$ can be extended  to $C=-1$, it is useful  to briefly review how  various  concepts involving high-energy hadron scattering have emerged in gauge/string duality. We also provide a short discussion on expectations for the $C=-1$ sector based on flat-space string scattering as well on the effect of confinement deformation.  For completeness, we will repeat  here some of the relevant discussions in \cite{Brower:2006ea}. Readers familiar with high energy hadronic collisions and the work in Ref.\cite{Brower:2006ea}  can move directly to Sec. \ref{sec:firstlook}.

\subsection{AdS Background and  Dual Pomeron} 

Conventional description of high-energy small-angle scattering in
QCD has two components --- a soft Pomeron  associated with exchanging  a tensor
glueball, and a hard BFKL Pomeron.
On the basis of gauge/string duality,  a coherent treatment
of the Pomeron was provided in Ref. \cite{Brower:2006ea}.   
In large-$N$ QCD-like theories, with beta functions that are vanishing or small in the ultraviolet, it has been shown  how the BFKL regime  and the classic Regge regime can be described simultaneously using curved-space string theory.

One important step in this development  involves the recognition 
 that in exclusive hadron scattering, the dual
string theory amplitudes, which in flat space are exponentially
suppressed at wide angle, instead give the power laws that are
expected in a gauge theory \cite{Polchinski:2001tt}.   
It has also been shown that at large $s$
and small $t$ the classic Regge form of the scattering amplitude is present in certain kinematic regimes~\cite{Polchinski:2001tt,Brower:2002er}. Next, deep inelastic scattering was studied \cite{DIS}.    At small $x$,  the
physics was found to be  similar to that of weak coupling, with a large growth
in the structure functions controlled by Regge-like  physics. 

Regge behavior  can also be approached from the IR  where confinement plays a central role.  For the $C=+1$ sector, it has been recognized earlier that,  with confinement deformation, transverse fluctuations of the metric tensor $G_{MN}$ in $AdS$ become massive, leading to  a tensor glueball~\cite{Brower:1999nj,Brower:2000rp,Constable:1999gb}. It was suggested in \cite{Brower:2000rp,Brower:1999tm} that, at finite $\lambda$,   exchanging such a tensor glueball, with  its associated Regge recurrences, would  lead to a Pomeron with an intercept below 2. That  is, a Pomeron can be considered as a  {\em Reggeized Massive Graviton.}

 The dual Pomeron was
subsequently identified as a well-defined feature of the curved-space string
theory~\cite{Brower:2006ea}.   The problem reduces to finding the spectrum of
a single $J$-plane Schr\"odinger operator,  or equivalently the
spectrum of the boost operator $M_{+-}$.
For ultraviolet-conformal
theories with confinement deformation,  the spectrum exhibits a set of Regge trajectories at
positive $t$, and a leading $J$-plane cut for negative $t$, the
cross-over point being model-dependent. (See Fig. \ref{fig:hardwallSpectrum}.)  For theories with
logarithmically-running couplings, one instead finds a discrete
spectrum of poles at all $t$, where the Regge trajectories at positive
$t$ continuously become a set of slowly-varying and closely-spaced
poles at negative $t$.  These  results agree with expectations for the
BFKL Pomeron at negative $t$, and with the expected glueball spectrum
at positive $t$, but provide a framework in which they are unified \cite{levintan,Bondarenko:2003xb}.

For conformally invariant gauge theories, the metric of the dual
string theory is a product, $AdS_5 \times W$,
\begin{equation}
\label{AdSWmetric}
ds^2 =\left( \frac{r^2} {R^2}\right) \eta_{\mu\nu} {dx^\mu dx^\nu} +\left(\frac{R^2} {r^2} \right)  {dr^2} + ds^2_W\ ,
\end{equation}
where $0 <r < \infty$.  We use $x^M$ for the ten-dimensional
coordinates, or $(x^\mu, r, \theta)$ with $\theta$ representing  the five
coordinates on $W$.  In this paper, we will ignore coordinates $\theta$ by concentrating on $AdS_5$ only.  For the dual to ${\cal N}=4$
supersymmetric Yang-Mills theory  the AdS radius
$R$ is
\begin{equation}
R^2 \equiv\sqrt{\lambda} \alpha'
= (4\pi g_{\rm string} N)^{1/2} \alpha' = 
(g_{\rm YM}^2 N)^{1/2} \alpha' \ ,
\end{equation}
and $W$ is a 5-sphere of this same radius.  We assume $\lambda \gg 1$, so that the spacetime
curvature is small on the string scale, and $g_{\rm string} \ll 1$ so
that we can use string perturbation theory.  It is often more useful to change variable from $r$ to $z=R^2/r$, so that the $AdS_5$ metric can be expressed as
\begin{equation}
\label{AdSmetricz}
ds^2 =\left( \frac{R^2} {z^2}\right)\left(  \eta_{\mu\nu} {dx^\mu dx^\nu} + {dz^2}\right)  + ds^2_W\ ,
\end{equation}
Another representation which we will make use of is 
\begin{equation}
\label{AdSmetricu}
ds^2 = e^{2u}   \eta_{\mu\nu} {dx^\mu dx^\nu} + R^2  {du ^2}  + ds^2_W\ ,
\end{equation}
where $z= R\; e^{-u}$.

\subsection{Flat-Space Expectation for $C=\pm 1$ Sectors}
\label{subsec:FlatSpace}

Let us begin by first establishing  some useful notations.  Consider
 two-body scattering, $a+b\rightarrow c+d$, in the near-forward limit of $s$   large,  and $t<0$   fixed, (with  all-incoming convention, $s = -(p_a + p_b)^2$  and  $t = -(p_a + p_c)^2$.) Another process related by crossing is $\bar c + b \rightarrow \bar a + d$, where $\bar a$ and $\bar c$ denote anti-particles of $a$ and $c$ respectively. Let us  denote amplitudes for these two processes by  $F$ and $\bar F$  and consider  the combinations $F^{\pm}\equiv  (\bar F \pm F)/2$, i.e., 
\bea
F_{\bar c b\rightarrow \bar a d } &\equiv&   \bar F = F^{+} + F^{-}\;, \nn
 F_{ab\rightarrow cd}   & \equiv &  F =  F^{+} - F^{-}\;.
\eea
However, since $F$ and $\bar F$ are also related by crossing, $s\leftrightarrow u$, $u=-(p_a+p_c)^2$, $F^{\pm}$, at fixed $t$, will be even and odd ($C=\pm 1$) in the difference variable: $\nu = (s-u)/2$.  At $s$ large and $t$ fixed, $\nu \simeq s$, and we shall therefore in what follows treat as if $F^{\pm}$ are even and odd in $s$.

For the most part, we will consider elastic scattering where $a=c$ and $b=d$. 
It follows that the average and the difference of total cross sections  at high energies are related to $C=\pm1$ amplitudes in the forward limit where $t=0$  via the optical theorem, i.e.,  
\bea
\left[ \sigma_T(\bar ab) + \sigma_T( ab) \right] & \sim  &  (2/s) \;{\rm Im}\; F^+\;,   \nn
 \left[ \sigma_T(\bar ab) - \sigma_T( ab) \right]  & \sim&    (2/s)\; {\rm Im} \; F^-\;.
\eea

Let us next  recall  how these amplitudes  are realized  for  string scattering in flat-space at high energy.  In a 10-dim flat-space, a crossing-even  string scattering amplitude at high energy takes on the form
\be
{\cal T}^{(+)}_{10}(s,  t) \to f^{(+)} (\alpha'  t) \left[ \frac{(-\alpha'  s)^{2 +\alpha'  t /2} +(\alpha'  s)^{2 +\alpha'  t /2}}{\sin \pi (2 +\alpha'  t /2)}\right]\;, \label{eq:10dflatspacePomeron}
\ee
where $ f^{(+)} (\alpha'  t)$ is process-dependent. We will re-derive this result in Sec. \ref{sec:Odderon} using OPE and we  will also introduce a Pomeron vertex operator which allows a direct generalization to the case of AdS background.  For now, we simply note that  Eq. (\ref{eq:10dflatspacePomeron})    corresponds to the exchange of a leading closed string trajectory
\be
\alpha_{+} (t) = 2 + \alpha'  t /2\;. \label{eq:gravitontraj}
\ee
That is, at $t=0$, for $C=+1$, one is  exchanging a massless spin-2 particle, i.e., the ubiquitous graviton. For a closed string theory~\footnote{In the
Regge limit we can ignore the fermionic modes, although strictly
speaking to avoid the tachyon and  to anticipate the AdS/CFT for ${\cal N}=4$ SUSY YM, we are actually
using the the critical  10-dim type-IIB superstring restricted
to the NS-NS sector. For a clear introduction to this topic, see ``A First Course in String Theory'', B. Zwiebach, Cambridge, 2004).}, massless modes 
in light-cone gauge are created by
a pair of left-moving and right-moving level-one oscillators $  a^\dagger_{1,I}\tilde a_{1,J}^\dagger$ from the NS-NS vacuum for type-IIB superstring:
\be
|I,J; k\rangle= a^\dagger_{1,I}\tilde a_{1,J}^\dagger |NS\rangle_L |NS\rangle_R |k\rangle \; .
\ee
In $D$ dimensions, there are $(D-2)^2$ such transverse modes, which can be grouped  into $D(D-3)/2$ traceless-symmetric components, $(D-2)(D-3)/2$ anti-symmetric components, and one component  for the trace. We shall denote representative states for each by
\be
|h\rangle = \sum_{I,J} h^{IJ}|I,J; k\rangle\quad ,\quad
|B\rangle =  \sum_{I,J}B^{IJ}|I,J; k\rangle\quad , \quad
|\phi\rangle =  \sum_{I,J}\eta^{IJ} |I,J; k\perp\rangle \; .
\label{eq:multiplet}
 \ee
with $h^{IJ}=h^{JI}$, $tr h=0$, $B^{IJ}=-B^{JI}$ and $k^2 = 0$. 
Since the 10-dim type-IIB superstring in the low-energy limit  becomes 10-dim super-gravity,these  modes  can be identified  with fluctuations of the metric $G_{MN}$, the anti-symmetric { Kalb-Ramond} background $B_{MN}$, and the dilaton, $\phi$, respectively.  For oriented strings, it can be shown that both the symmetric tensor and the scalar contribute to $C=+1$ and the anti-symmetric tensor contributes to $C=-1$.  

It is worth pointing out that, for $D=4$, there are only two independent traceless-transverse metric fluctuations, $D(D-3)/2=4(4-2)/2=2$, which corresponds precisely to helicity $\lambda = \pm 2$ for the massless graviton. However, in AdS/CFT, we will be dealing with $AdS_5$, where $D=5$. This leads to $5(5-3)/2=5$ independent  modes, which is precisely what is necessary for  turning the graviton massive on $AdS_5$, i.e.,  a tensor glueball, as demonstrated in Refs. \cite{Brower:1999nj,Brower:2000rp,Brower:1999tm}.  This is also summarized in Table~\ref{tab:classIIB}.

For the $C=-1$ sector, the amplitude  behaves as 
\be
{\cal T}^{(-)}_{10}(s,  t) \to f^{(-)}(\alpha'  t)\left[ \frac{(-\alpha'  s)^{1 +\alpha'  t /2} - (\alpha'  s)^{1 +\alpha'  t /2}}{\sin \pi (1 +\alpha'  t /2)}\right] \ , \label{eq:10dflatspaceOdderon}
\ee
corresponding to the exchange of a closed string trajectory
\be
\alpha_{-}(t) = 1 + \alpha'  t /2\;.\label{eq:KRlineartraj}
\ee
This can again be understood as exchanging  states lying on a leading $C=-1$ trajectory. Consider   the  state $|B\rangle $  obtained by applying $\sum_{IJ} B_{IJ}a^\dagger_{1,I}\tilde a^\dagger_{1,J}$, where $B_{IJ}=-B_{JI}$,  to the NS-NS vacuum. One can again check explicitly that  these are massless, i.e.,  $k^2=0$. It is also useful to note  that,  for $D=4$, there is only one independent component for the anti-symmetric tensor, i.e.,  $(D-2)(D-3)/2=1$, which 
leads to a zero-helicity state. The  ``helicity $\pm 1$'' components we are seeking are  pure gauge and  unphysical. That is, the leading component must  de-couple by the absence of a pole in $f^{(-)}(t)$ at $t=0$.    However, since we will eventually be working in $AdS_5$, the dimension is one higher, and the number of independent modes is $(5-2)(5-3)/2= 3$, so that the desired components for spin-1 survive in this limit \cite{Brower:2000rp}.

As we have also explained in Ref. \cite{Brower:2006ea}, in the conformal limit,  gauge/string duality turns the flat-space graviton  trajectory, (\ref{eq:gravitontraj}),  into  a $J$-plane  branch cut.  In Secs. \ref{sec:firstlook} and \ref{sec:Odderon}, we explain how the corresponding $C=-1$  flat-space exchange, (\ref{eq:KRlineartraj}),   gets modified  in the AdS-background, leading to (\ref{eq:dualconformalOdderon}).  The strong coupling conformal Odderon is again fixed cuts in the $J$-plane.

\subsection{Confinement Deformation} 
\label{sec:confinement}

We are ultimately interested in gauge theories that are near conformally-invariant in the UV, (broken only by logs due to asymptotic freedom), but with conformal invariance strongly broken in the IR, resulting in a mass gap and confinement.  If confinement sets in at a scale $\Lambda$ in the gauge theory, this leads to a change in the metric away from $AdS_5\times W$ in the region near $r \sim \Lambda R^2 \equiv r_0 $. Roughly speaking, this means that the dual string metric is of the $AdS$ form but with a lower cutoff on the coordinate $r$, so that $r_0 < r< \infty$.  More precisely, for QCD, the product structure breaks down in the infrared.  The precise metric when $r\sim r_0$ depends on the details of the conformal symmetry breaking.  Most of the physics that we will study takes place in the conformal region where the metric is the approximate $AdS$ product~(\ref{AdSWmetric}). Even here we might generalize to geometries that evolve slowly with $r$, as in the running coupling example studied in \cite{Brower:2006ea}.  In terms of $z$, the infrared cutoff $z_0= R^2/r_0$ becomes an upper cutoff. Similarly, for $u$, there will be a lower infrared cutoff $u_0= -\log (z_0/R)= \log (r_0/R)$. Alternatively, we can re-define $u$ by $z=z_0e^{-u}$, so that $u_0=0$.

In kinematic regimes where confinement plays no role, it is sufficient to adopt the AdS metric, appropriate for  conformally-invariant theories.  To  consider the generic effects of
confinement, it is sufficient to modify the metric near $r\simeq r_0$, while keeping the ultraviolet nearly conformal.    A simple
example of such a theory is the ${\cal N}=1^*$ model studied in
\cite{nonestar}.  This discussion is by necessity less precise than
the previous ones, simply because there is model-dependence in the
confining region.   Typically the space is cut off, or rounded
off, in some natural way at $r=r_{0}$. This allows one  to make as many
model-independent remarks as possible, and examine where
model-dependence is to be found. This
leads to a theory with a discrete hadron spectrum, with mass
splittings of order $\Lambda$ among hadrons of spin $\leq 2$.  The
theory will also have confining flux tubes (assuming these are stable)
with tension $1/\alpha'_{0} = 2\sqrt{\lambda}\Lambda^2$; the same scale
sets the slope of the Regge trajectories for the high-spin hadrons of
the theory.  Note the separation of the two energy scales, $\Lambda$ and $\alpha_0^{-1/2}$, by a factor
of $\lambda^{1/4}$; this is an important feature of the
large-$\lambda$ regime.

\begin{figure}[bthp]
\begin{center}
\includegraphics[width = 3.5 in]{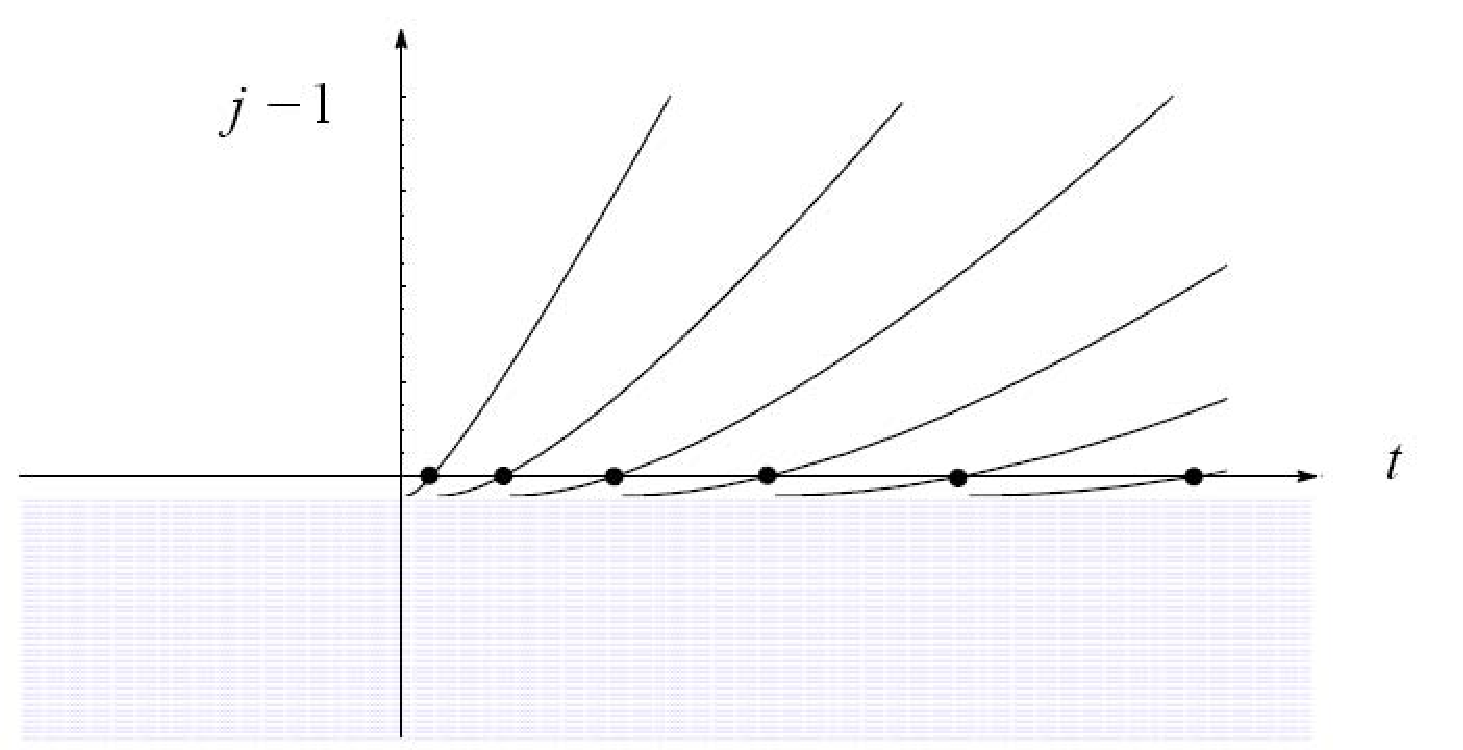}
\end{center}
\caption{The $J$-plane structure  for the $C=-1$ sector is schematically represented  by the spectrum for hard-wall model. As $t$ decreases, trajectories move under the fixed-cut at $j=j^{(-)}_0$ and disappear from the physical sheet. The transition region from moving-pole to fixed cut is model-dependent. At sufficiently large $t$
  each trajectory attains a fixed slope, corresponding to the tension
  of the model's confining flux tubes.  This $J$-plane structure for the $C=-1$ sector is similar to that for the $C=+1$ sector, as depicted in Fig. 8 of Ref. \cite{Brower:2006ea}. }
\label{fig:hardwallSpectrum}
\end{figure}

Since the metric is changed near $r_{0}$, the associated Schr\"odinger problem approaches
the conformal  limit only for $ r\gg r_{0}$.  However, for $-t\gg
\Lambda^2$, it has been shown in \cite{Brower:2006ea} that the effective potential  is insensitive to the region
near $r_0$.  This is consistent with the expectation in the QCD
literature that the BFKL calculation is infrared-safe for
large negative $t$, while the effects of confinement become important as
$t\to 0^-$, and for  $t>0$. 
For illustrative purpose, it is instructive to work
with  a ``hard-wall'' toy model.  The main advantage of the hard-wall model is that it can be treated
analytically.  Since the
metric is still $AdS_5\times W$, we have the same quantum mechanics
problem to solve as in the conformal case  except for a cutoff on the space at $r_0$. While this model is 
not a fully consistent theory, it does capture key features
of confining theories with string theoretic dual
descriptions.  For instance, the resulting $J$-plane spectrum which  exhibits a set of Regge trajectories at positive $t$ and the presence of an BFKL cut can be schematically represented by Fig.  \ref{fig:hardwallSpectrum}. 
We will return to these features for both $C=\pm1$ sectors in Sec. \ref{sec:spectrum}. 

\newpage

\section{Conformal Odderon Propagator in Target Space}
\label{sec:firstlook}

In this section, we begin by first providing a simplified discussion for both  the $C=\pm 1$ sectors based on the ``ultra-local'' approximation.  Just like the case for the Pomeron  discussed in \cite{Brower:2006ea}, this  leads to a  ``red-shifted'' Odderon. This effective Odderon is  linear for $t>0$ and is a constant near $j=1$ for $t<0$, with a kink at $t=0$. We next introduce diffusion, moving from the flat-space scattering to AdS.  We will  restrict ourselves  to  the conformal limit.  By turning the problem into an equivalent Schr\"odinger   problem, the $J$-plane singularity follows from a standard spectral analysis. In particular, in the conformal limit, the spectrum consists of a continuum, corresponding to a $J$-plane branch-cut, (\ref{eq:dualconformalOdderon}).

\subsection{Ultra-Local Approximation}

 To  see how Regge behavior differs in a curved-space  from  a flat-space,  let us begin by providing an heuristic treatment.   In the case of $AdS$, let us  first assume that   scattering takes place ``locally'', i.e.,  only when all particles are located at the same  $r$.  The local inertial quantities
are
\begin{equation}
\tilde s = \frac{R^2}{r^2} s\ , \quad \tilde t = \frac{R^2}{r^2} t\ ,
\end{equation}
and, with  $r$ fixed,  the ten-dimensional scattering process remains  in the Regge regime when $s$ is sufficiently large.  Thus at fixed $r$ we have, instead of (\ref{eq:10dflatspacePomeron}) and (\ref{eq:10dflatspaceOdderon}), 
\be
{\cal T}^{(\pm)}_{10}(\tilde s, \tilde t) \sim  f^{(\pm)}(\alpha' \tilde t) (\alpha' \tilde s)^{\alpha_{\pm}(0)+\alpha' \tilde t /2} \sim s^{\alpha_{\pm}(0)+\alpha'_{eff}(r)   t /2},  \label{eq:regten}
\ee
where  $\alpha_{+}(0)=2$, $\alpha_-(0)=1$, and 
\be 
\alpha'_{\rm eff}(r) = \frac{R^2
  \alpha'}{r^2} \; .
\label{alfeff} 
\ee 
with $f^{(\pm)}$  process-dependent functions. We have also left  out wave functions,  $\{\psi_i(r) \}$,  for four external hadrons.    It follows that one arrives at scattering in 4-dim by   summing  over the $AdS$ radius, i.e., 
\begin{equation}
{\cal T}^{(\pm)}_4(s,t) \sim \int dr \, \sqrt{-G}\, 
{\cal T}^{(\pm)}_{10}(\tilde s, \tilde t)  \ , \label{ramp}
\end{equation}
leading to a  4-dim Pomeron and Odderon respectively by superpositions.  By examining the exponent of $s$, we see that the intercepts
 at $t= 0$ 
remain at 2 and   1 respectively,  just as in flat spacetime.  We also see that the slope of $t$, $\alpha'_{eff}$, depends on $r$.
 It is as though, in this ``ultralocal'' approximation, each  pair of 
five-dimensional Pomeron and Odderon  gives rise to a  continuum of four-dimensional
Pomerons and Odderons, one pair for each value of $r$ and each with a slope $\sim 1/r^{2}$.~\footnote{
The notion of a tension depending on a fifth dimension
dates to \cite{Polyakov:1997tj}. 
The idea of superposing many five-dimensional Odderons is conceptually
anticipated in the work of \cite{Brower:2002er}.}

\begin{figure}[bthp]
\begin{center}
\includegraphics[width = 3 in]{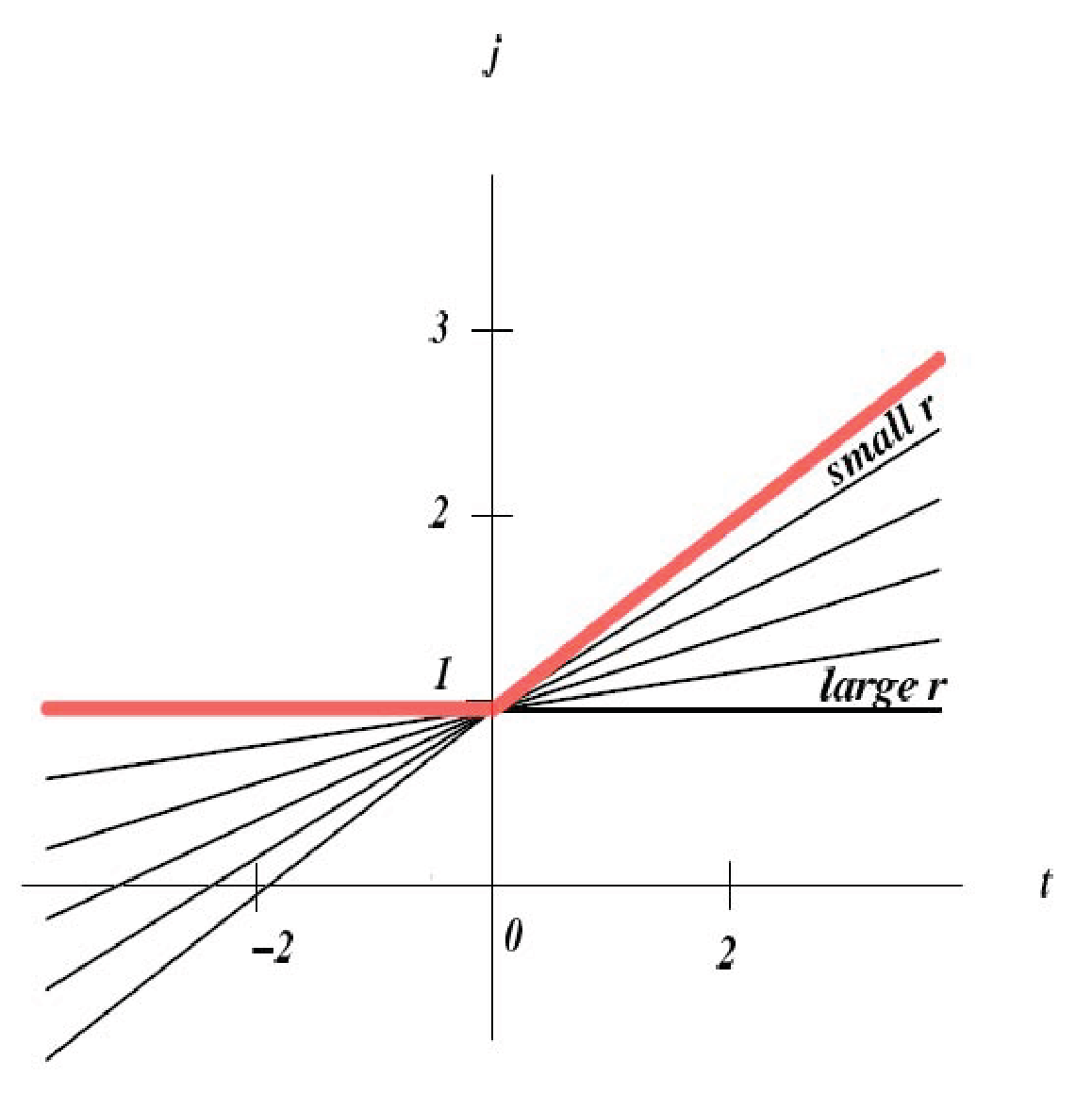}
\end{center}
\caption{ In the ultralocal approximation, the leading singularity for $C=-1$ (the singularity with largest $j$ at fixed
$t$) is the usual linear Regge trajectory for $t>0$ but is a constant at $j=1$ 
 for $t<0$.   The ultralocal picture for $C=+1$ is identical, except for the intercept at $t=0$ shifted up to $j=2$, as  illustrated by Fig. 1 of Ref. \cite{Brower:2006ea}. }
\label{fig:ultralocal}
\end{figure}

At large $s$, the highest trajectory will dominate.  Recall, for the Pomeron, at 
positive $t$, this would be the one at the minimum value $r_{0}$:
\be
\alpha^{(+)}_{eff}(t)\simeq 2 + \alpha'_{eff}(t>0) \; t \; , \quad {\rm and} \quad \alpha'_{\rm eff} (t > 0) = \frac{R^2 \alpha'}{ r_0^2} \equiv \alpha'_0
\ . \label{alf0} 
\ee
For negative $t$, it would be the trajectories at large $r$.  The
wavefunctions in the superposition~(\ref{ramp}) make the integral
converge at large $r$, so at any given $s$ the dominant $r$ is finite,
but as $s$ increases the dominant $r$ moves slowly toward $\infty$  and
so we have 
\be 
\alpha'_{\rm eff} (t < 0) \simeq 0 \ .  
\ee 

Same analysis also applies to the Odderon.  For positive $t$, the effective Odderon
sits at small $r$ and so its properties are determined by the
confining dynamics.  At negative $t$, the effective ``spin'' of an exchanged Odderon is
\be
\alpha^{(-)}_{\rm eff} (t) \simeq  1 + \frac{ 1}{2\sqrt \lambda} \frac{R^4}{r^2} t
\ee
The largest value  sits at
large  $r$ and so is effectively a very small object, analogous to the
tiny (and therefore perturbative) three-gluon Odderon.  

This is illustrated schematically in Fig.~\ref{fig:ultralocal}.
We see that, under this approximation, the
dominant  Odderon trajectory has a kink at $t = 0$, as noted in \cite{Brower:2006ea}  for the Pomeron.

\subsection{Diffusion in AdS}
\label{sec:diffusion}
However, the ultralocal approximation ignores diffusion in $AdS$.  Regge behavior is intrinsically non-local in the transverse space.  For flat-space scattering in 4-dimension, the transverse space is the 2-dimensional  impact space, $\vec b$. In the Regge limit of $s$ large and  $t<0$, the momentum transfer is transverse. Going to the $\vec b$-space, 
\be
t   \to   \nabla_b^2\;, 
\ee
and the flat-space Regge propagator, for both $C=\pm 1 $ sectors,  is nothing but a diffusion kernel
\be
  \langle \; \vec b \;|\;(\alpha' s)^{\alpha_{\pm}(0)+\alpha' t/2} \;| \; \vec b' \;\rangle \to    (\alpha' s)^{\alpha_{\pm}(0) } \;   \frac {e^{ - (\vec b-\vec b')^2/(2{\alpha'}^2 \tau)}}{ \tau^{(D-2)/2}}
\ee
with $\tau=\log (\alpha' s)$, $\alpha_+(0)=2$ and $\alpha_-(0)=1$. 
Therefore, in moving to a
ten-dimensional momentum transfer $\tilde t$,   we must keep a term previously
dropped, coming from the momentum transfer in the six transverse directions. 

Let us first focus on $C=+1$ sector. This extra term leads to diffusion in  extra-directions, i.e., 
\be
\alpha'  \tilde t \to  {\alpha'} \Delta_{P} \equiv
\frac{\alpha' R^2 }{r^2} \nabla_b^2+ {\alpha'} \Delta_{\perp P}\ . 
\label{kinemt}
\ee
The transverse Laplacian is proportional to $R^{-2}$, so that the added term
is indeed of order ${\alpha'}/{R^2} = 1/\sqrt{\lambda}$.  The Laplacian acts
in the $t$-channel, on the product of the wavefunctions of states 1 and 2 (or
3 and 4).  The subscript $P$ indicates that we must use the appropriate
curved spacetime Laplacian for the Pomeron being exchanged in the
$t$-channel  \cite{Brower:2006ea}.  We  see
that $\tilde s^{\alpha' \tilde t /2}\sim e^{({\alpha'} \Delta_{P}/2)  \tau    }$ is now a diffusion operator in all
eight transverse dimensions, not just the Minkowski
directions.  By ignoring  the $\theta$ variables, $ \Delta_{\perp P}$ involves only the AdS radius r.

To obtain the $C=+1$ Regge exponents we will have to diagonalize the differential
operator~(\ref{kinemt}). Instead of an  ``ultralocal''
Pomeron, we will
have a
more normal spectral problem.\footnote{The structure that we find in AdS/CFT, where the trajectories
are given by the eigenvalues of an effective Hamiltonian, 
closely parallels to that found by BFKL in perturbation theory.}  Using a Mellin transform,  $\int_0^{\infty} d\tilde s \; {\tilde s}^{-j-1}$,  the Regge propagator can be expressed as 
\be
\tilde s^{ 2+ \alpha' \tilde t /2 }
 =   \int \frac{d j}{2\pi i} \; \frac{{\tilde s}^j }   { j-  2 -\alpha' \Delta_P /2 }
\ee
As shown in \cite{Brower:2006ea},  $\Delta_P \simeq  \Delta_j  $, the tensorial Laplacian, and it is related to the scalar $AdS$ Laplacian $\Delta_0$ by $\Delta_j  = r^j \; \Delta_0\; r^{-j}$.  Making use of the EOM for  traceless-transverse metric fluctuations,  $\Delta_2 h_{MN}=0$,  it is sufficient to replace $\Delta_j $ by $\Delta_2$  for   determining  the first strong coupling
correction, $O(1/\sqrt{\lambda})$, to the Pomeron intercept,  (\ref{eq:dualconformalPomeron}).  It is also convenient to introduced a $J$-plane $C=+1$ propagator $G^{(+)}(j)$.   
\be
G^{(+)}(j) = \frac{1}{ j-  2 -\alpha' \Delta_2 /2}  \label{eq:pomeronpropagator}
\ee
In Ref. \cite{Brower:2006ea}, one finds that this leads to a $J$-plane cut. We will return to this  shortly. 

A similar analysis can next be carried out for the $C=-1$ sector. In moving to AdS, we simply replace the  Regge kernel by 
\be
\tilde s^{ 1+ \alpha' \tilde t /2 } =\int \frac{d j}{2\pi i} \; {\tilde s}^j \;    G^{(-)}(j) =   \int \frac{d j}{2\pi i} \; \frac{{\tilde s}^j }   { j-  1 -\alpha' \Delta_O /2 }
\ee
where we have again introduced a $J$-plane $C=-1$  propagator. The operator $\Delta_O(j)$ can be fixed by examining  the EOM at $j=1$ for the associated super-gravity fluctuations responsible for this exchange, i.e., the anti-symmetric Kalb-Ramond field, $B_{MN}$.  There are two classes of EOM's:
\be
(\square_{Maxwell} - (k+4)^2)B^{(1)}_{IJ}=0\; , \quad (\square_{Maxwell} - k^2)B^{(2)}_{IJ}=0
\ee
$k=0, 1,2,\cdots$, where $\square_{Maxwell}$ stands for the Maxwell operator  defined in Eq.~\ref{eq:Maxwell}.   However, for $k\neq 0$, there would be associated fluctuations on W, which we ignore, thus leaving two solutions, both for $k=0$.  We also normalize the $\square_{Maxwell}$ operator so that its component in flat transverse directions is $(R^4 /r^2) \nabla_b^2$. For both cases, we have
\be
G^{(-)}(j) =  \frac{1}{ j-1 -(\alpha'/2R^2) (\square_{Maxwell} - m^2_{AdS,i}) }   \label{eq:Odderonpropagator}
\ee
where the AdS mass-squared are~\footnote{In our normalization, these are dimensionless. Formally, the second  solution can be gauged away. However, we leave it here for now and  will return to have a more careful discussion  later. }
\be
m^2_{AdS,1}=16\; , \quad m^2_{AdS,2} = 0\; .
\ee

We are now in position to  demonstrate  that (\ref{eq:pomeronpropagator}) and (\ref{eq:Odderonpropagator})   lead  to diffusion in AdS for $C=+1$ and $C=-1$ sectors respectively. We shall treat both cases in parallel. It is  convenient to switch to coordinate $z= R^2/r$ for the $AdS$ coordinate.  Denote matrix elements by $G^{(\pm)}( z,z,'; j,t) =  \langle z;t \; |  G^{(\pm)}(j)  |\; z';t\rangle $, 
we adopt a normalization so that $(zz')^2G^{(\pm)}$ each satisfies a standard $AdS_5$  scalar   equation.
  Indeed, at $j=2$ and $j=1$ respectively, each reduces precisely to the equation for an $AdS_5$ scalar propagator, up to a constant factor $2\sqrt \lambda$. 
  It is actually  simpler to deal with $G^{(\pm)}$ directly, and one finds 
\be
 (1/2\sqrt \lambda) \left  \{   -z \dd_z z \dd_z + {z^2} t   + m^2_{\pm}(j) \right\}  G^{(\pm)}( z,z'; j,t)  =z \; \delta(z-z') \label{eq:DE2}
\ee 
where   $m^2_+(j)= 2\sqrt\lambda(j-2) + 4$   and $m^2_-(j)= 2\sqrt\lambda(j-1) + m^2_{AdS,i}$.  $G^{(\pm)}$ can now be found by a standard spectral analysis.

 Recall that a diffusion operator resembles a Schr\"odinger operator in
imaginary time, and the desired diagonalization can be similarly treated.  To turn this into a conventional Schr\"odinger-like equation,  introduce a new variable $u$, where $z=z_0 e^{-u}$ and $z_0$ arbitrary. For simplicity, we will set $z_0=1$ in what follows. Aside from the $J$-dependent term $m^2_\pm(j)$, we  need only to solve a standard Schr\"odinger eigenvalue problem, e.g.,
\begin{equation}\label{SchrodingerEq}
H^{(\pm)}\Psi_E(u)=[-\partial_u^2 + V(u,t)]\Psi_E(u) = E\Psi_E(u)
\end{equation}
with $V(u,t)= - t e^{-2u} $.   
We can simplify the problem further by considering $t=0$ first. This has the same eigenstates as a
free particle.  The spectral representation~\footnote{A cautionary note:  $\nu$  used in Refs.~\cite{Brower:2006ea,Brower:2007xg}     is  2 times the $ \nu$  used here,  and this has the effect that the
  diffusion constant $\cal D$ here is four times that   in Refs.~\cite{Brower:2006ea,Brower:2007xg}.} for $G^{\pm}$ at $t=0$ is
\be
G^{(\pm)}(z,z'; j,0) = \frac{1}{\sqrt \lambda} \int_{-\infty}^\infty  d\nu \frac {e^{ 2i \nu (u-u')}}{ j - j_0^{\pm}+{\cal D} \nu^2}   \label{eq:Gpmt0}
\ee
with $ j^{(\pm)}_0$ given by (\ref{eq:dualconformalPomeron}) and  (\ref{eq:dualconformalOdderon}), and ${\cal D}= 2/ \sqrt\lambda$.

Upon taking an inverse Mellin transform, it  follows that
\be
 \langle z ; 0| \tilde s^{ \alpha_{\pm}(0) + \alpha' \tilde t /2 } | z'; 0 \rangle = e^{j^{(\pm)}_0\tau} \; \frac {e^{ - 2 (u-u')^2/ {\cal D} \tau}}{ \sqrt{ \pi {\cal D}  \tau}}
\ee
where we have $\tau=\log \tilde s$.
This is the
same type  of behavior  found by BFKL in perturbative context. Here, it corresponds to  diffusion in AdS. In particular, we can
identify $j^{(\pm )}_0$ as the strong coupling limit of the Pomeron and Odderon intercepts respectively.  Also from (\ref{eq:Gpmt0}), one can verify that  there is a branch point  in the $J$-plane ending at $j^{(\pm)}_0$, as summarized in Table \ref{tab:intercepts}, for both $C=\pm 1$.

For $t\neq 0$, the eigenvalue problem can again be solved in terms of Bessel functions while the spectrum remains unchanged. Instead of (\ref{eq:Gpmt0}), one has
\be
G^{(\pm)}(z,z'; j,t) = \frac{2}{\sqrt \lambda \pi^2} \int_{-\infty}^\infty  d\nu \; \nu \sinh2 \pi \nu  \frac {K_{2i\nu}(|t|^{1/2} e^{-u}) K_{-2i\nu} (|t|^{1/2} e^{-u'})}{ j - j_0^{\pm} + {\cal D} \nu^2} 
\ee
Therefore, the spectrum in the $J$-plane  remains a fixed cut at $j=j^{(\pm)}_0$.  

It is also useful to explore the conformal invariance as the  isometry of  transverse $AdS_3$.  As shown in \cite{Brower:2007xg}, upon taking a two-dimensional Fourier transform with respect to $q_\perp$,  where  $t=-q_\perp^2$, one finds  that $G^{(\pm)}(z,x^\perp,z',x'^{\perp}; j)$ can be expressed simply as
\be
G^{(\pm)}(z,x^\perp,z',x'^{\perp}; j)  = \frac{1}{4\pi zz' } \frac{e^{ (2 -
\Delta^{(\pm)}(j))\xi}}{  \sinh \xi} \; .
\label{eq:adschordal}
\ee
The variable $\xi$ is related to the $AdS_3$ chordal distance  
 \be\label{eq:vdefn}
v= \frac{(x^\perp-x'^{\perp})^2+(z-z')^2}{2 zz'}
\ee
by $\cosh \xi =1 +v$,  and 
\be
\Delta^{(\pm)} (j) =  2 + \sqrt {2} \;
\lambda^{1/4} \sqrt{(j-j^{(\pm)}_0) }
\label{eq:Delta4}
\ee
is a  $J$-dependent effective  $AdS_5$ conformal dimension~\footnote{For $\lambda $ large, $\Delta^{(+)} (j) \simeq \sqrt 2 \; \lambda^{1/4} \sqrt {j-2} + O(1)$ for $j>2$. This feature has also been emphasized recently by Hofman and Maldacena \cite{Hofman:2008ar} in a related context. For Regge, we are interested in the region $\Delta\simeq 2$ where $j_0^{(+)} < j  <2$. Note that  $\Delta^{(+)}(2)=4$ for all $\lambda$, due to energy-momentum conservation. Will return to this point in Sec. \ref{sec:general}.}.

For completeness, we note that,  for both $C=+1$ and $C=-1$, 
it is  useful to  introduce Pomeron and Odderon kernels in a mixed-representation,
\be
{\cal K}^{(\pm)} (s, z,x^\perp, z',x'^\perp) \sim \left(\frac{(zz')^2}{R^4}\right) \int \frac{dj}{2\pi i}   \;     \left[ \frac{(-\tilde s)^j \pm (\tilde s)^j}{\sin \pi j} \right] G^{(\pm)} (z,x^\perp,z',x'^\perp;j)  \;. \label{eq:PomeronOdderonkernel}
\ee
  To obtain scattering amplitudes, we simply fold these kernels with external wave functions. Eq. (\ref{eq:PomeronOdderonkernel}) also serves as the starting point for eikonalization; we will return to this point in Sec. \ref{sec:conclusion}.

\subsection{Conformal Geometry at High Energies}
\label{sec:geo}

We now turn to a brief discussion on the conformal invariance of the $C=\pm 1$ kernels, $G^{(\pm)}(j)$. As pointed out in Ref.~\cite{Brower:2007xg} the strong coupling AdS Pomeron ($C= +1$) kernel is almost completely determined by conformal symmetry or more precisely the subgroup of the conformal group that commutes with the boost operator $M_{+-}$. In the dual theory this  subgroup corresponds to the $SL(2,C)$ isometries of the Euclidean $AdS_3$ submanifold transverse to the lightcone co-ordinates: $x^0 \pm x^3$,  (see Appendix~\ref{app:conformal} for details.) As such it plays the same role as the little group which commutes with the energy operator $P_0$.  Here we extend this observation to the Odderon ($C = -1$) kernel.

Consider  a general   $n$-particle scattering amplitude, $A(p_1,p_2, \cdots p_n)$.  Regge exchange corresponds to having a large rapidity gap separating the $n$ particles into two sets: the right movers and left movers, with large $p^+_r = (p^0_r+ p^3_r)/\sqrt{2}$ and large $p^-_\ell = (p^0_\ell - p^3_\ell)/\sqrt{2}$ momenta respectively \cite{Brower:2006ea,Brower:2007xg}.  The $C=\pm 1$ kernels are obtained by applying this limit to the leading diagrams, in the $1/N_c$ expansion, that carries respective vacuum quantum numbers in the $t$-channel.
The rapidity gaps, $\ln(p^+_r p^-_\ell)$, between any right- and
left-moving particles are all $O(\log s)$, i.e., a large Lorentz
boost, $\exp[y M_{+-}]$, with $y\sim \log s$, is required to switch
from the frame comoving with the left movers to the frame comoving
with the right movers. Clearly, the large $y$ behavior is controlled by the leading spectrum of $M_{+-}$.

Since the $J$-plane is conjugate to rapidity, upon taking a Mellin transform,   the spectrum of the boost generator $M_{+-}$ can be found  by examining the resolvent, $(j-M_{+-})^{-1}$. This resolvent is nothing but the Regge propagator for $C=\pm 1$ respectively.
To evaluate  this propagator, we can go to a basis where $M_{+-}$ is diagonal. Since  $M_{+-}$ commutes with all the generators
of $SL(2,C)$, one finds  
 that  the propagator $(j-M_{+-})^{-1}$  can be  expressed in terms of $SL(2,C)$ Casimirs. 
 
From a more intuitive perspective,   one recognizes that  high energy small-angle scattering can be  separated into
longitudinal and transverse, relative to the momentum
direction of the incoming particles.  The transverse subspace of
$AdS_5$ is $AdS_3$.  It is therefore not surprising to find  that 
the co-ordinate representation for both the $C=\pm 1$ Regge  propagators 
can be expressed as bulk-to-bulk scalar propagators in the
transverse Euclidean $AdS_3$ with $SL(2,C)$ isometries. Very likely this
is a generic property  in all conformal gauge theories.

 In Appendix B, we briefly review how $SL(2,C)$ generator algebra can be realized on  $AdS_3$. In general,
unitary representations of $SL(2,C)$ are labeled by $h = i
\nu + (1+n)/2 $, and $\bar h = i \nu +(1-n)/2$, which are the eigenvalues 
for the highest-weight state of $J_0$ and $\bar
J_0$.  The principal series is given by real $\nu$ and integer $n$.    
For our leading order strong coupling Pomeron and Odderon, one finds that  $J^2 = \bar
J^2= -(1/4 +\nu^2)$;     we are thus
restricted to $n = h - \bar h = 0$ and are insensitive to rotations in
the impact parameter plane by $M_{12}$.   

Let us focus here on the $C=-1$ sector.  To leading order in strong
coupling, the boost operator can be  expressed as  $M_{+-} = 1 -
H^{(-)}_{+-}/(2 \sqrt{\lambda}) + O(1/\lambda)$, with $H^{(-)}_{+-}$ given by the DE obeyed by the propagator $G^{(-)}$, or equivalently,  $zz'G^{(-)}$, i.e., Eq. (\ref{eq:G3}).   From (\ref{eq:J2}), we find
\be
H^{(-)}_{+-}   = -z^3\dd_z z^{-1} \dd_z   - z^2  \nabla^2_{x_\perp} + m^2_{AdS} -1= - 2 J^2 - 2 \bar J^2+m^2_{AdS} -1 \; .
\label{eq:strongH}
\ee
After a similarity transformation,   $ H^{(-)}_{+-} $  can be diagonalized by a Fourier transform, with eigenvalue $-\infty< \nu<\infty$ and eigenfunction
\be
 Y_{\nu}(v)=  \frac{1}{4\pi } \frac{e^{ 2i\nu\xi}}{  \sinh \xi} \;.\label{eq:eigenfunction}
\ee
  It follows that 
the  boost operator $M_{+-}$ is diagonal in the $\nu$-basis   and, in the strong coupling,
\be
M_{+-}\rightarrow j^{(-)}(\nu)= j^{(-)}_0 -  {\cal D} \nu^2 + 0(\nu^4)  \; .
\ee
with  $j^{(-)}_0=1-m^2_{AdS}/\sqrt\lambda$ and  ${\cal D} =
2/\sqrt\lambda$.

We are now in the position to put everything together, for both $C=+1$ and $C=-1$. 
It is useful to  introduce Pomeron and Odderon propagators in a mixed-representation,
\be
 \left(\frac{1}{zz'}\right)  \int \frac{d\nu}{ \pi} \; \frac{ 1} { j-j^{(\pm)}_0 + {\cal D} \nu^2 }   {Y}_{\nu}(v)  \;.
\ee
Closing the $\nu$-contour, one obtains  $G^{(\pm)} (z,x^\perp,z',x'^\perp;j)$, given by Eq. (\ref{eq:adschordal}).

It is interesting to note that this strong coupling Hamiltonian structure, (\ref{eq:strongH}),  is similar to the weak
coupling one-loop $n_g$ gluon BFKL spin chain operator in the large $N_c$
limit.  Here the boost operator is approximated by $M_{+-} = 1 -
(\alpha N_c/\pi) H_{\rm BFKL}$, where 
\be
H_{\rm BFKL}
= \frac{1}{4}\sum^{n_g}_{i=1}[{\cal H}(J^2_{i,i+1}) + {\cal H}(\bar
J^2_{i,i+1})],   \label{eq:BFKLleadingH}
\ee
is a sum over two-body operators  with holomorphic and
anti-holomorphic functions of the Casimir. The Yang-Mills coupling is
defined as $\alpha= g^2_{YM}/4\pi$. Even numbers of gluons ($n_g$)
contribute to the BFKL Pomeron with charge conjugations $C=+1$ and
odd number of gluons to the weak coupling  Odderon~\cite{Kwiecinski:1980wb,Janik:1998xj,Braun:1998fs,Bartels:1999yt} with charge
conjugations $C=-1$.  More discussion on  the relation between the strong-coupling and
 weak-coupling limits  can be found in \cite{Brower:2007xg}.

\newpage
\section{Odderon Vertex  on the World-Sheet}

\label{sec:Odderon}

It has been demonstrated  in Ref. \cite{Brower:2006ea} that Regge behavior for string scattering in flat-space follows from an OPE on world-sheet. Since the
flat space (super) string is integrable with an explicit closed form
for the amplitude, it is possible to define in closed form the on-shell Reggeon vertex.
For example the on-shell Pomeron vertex is given by
\be
{\cal V}_P^{\pm} = (2\dd X^{\pm}\bar \dd X^{\pm}/\alpha')^{1+\alpha' t/4} e^{\mp i k\cdot X} \label{eq:Pomeronvertexop} \; .
\ee
(See \cite{Hofman:2008ar} for a related discussion.)  This operator
(and a similar Reggeized gauge field operator for the open string) enables
a very elegant direct evaluation of Regge and multi-Regge
amplitudes reproducing all the results in the early literature,
for example as reviewed  in Ref.~\cite{Brower:1974yv} for the open string.

In $AdS_5 \times S^5$, without a closed form solution for amplitudes, no closed form particle or Regge vertex operator has been constructed. However Ref.~\cite{Brower:2006ea} did succeed in finding the Pomeron vertex  to leading order in the strong coupling approximation.  A key step involves in  identifying the Pomeron
source as a generalized  (1,1) conformal  worldsheet vertex operator, satisfying the on-shell conditions: $L_0=\bar L_0= 1$. Here we will generalized this analysis for the Odderon.

We begin by treating the bosonic sector of the supergravition multiplet in flat space introduce in Sec.~\ref{subsec:FlatSpace}, Eq.~(\ref{eq:multiplet}). The natural guess is take the
high energy limit for a more general Reggeon operator insertion,
\be
{\cal V}(T) = (T_{MN} \dd X^{M}\bar \dd X^{N}/\alpha')^{1+\alpha' t/4} e^{\mp i k\cdot X} \; ,
\ee 
where one can  expand ${\cal V}(T)$ into a leading symmetric traceless ($h_{MN}$), antisymmetric ($B_{MN}$) and the trace ($\phi\; \eta_{MN}$) contributions, appropriate  for the Pomeron, Odderon and dilaton respectively. Generalization to the entire supergravity multiplet including the Fermionic fields is straight forward in principle but truncating to bosonic degrees of freedom has the advantage of being able drop all world sheet Grassmann variable from the outset.

\subsection{Pomeron/Odderon Vertex Operator in Flatland}

One of the key observation made in \cite{Brower:2006ea} is the recognition that  Regge behavior in flat-space scattering  follows also directly from the world sheet OPE.  For instance, for the standard bosonic closed string scattering involving 4 external tachyons,  the amplitude is given by
\bea
A_{4}&=& \intop d^{2}w\; \langle e^{ip_1\cdot X(0)} e^{ip_2\cdot X(w,\bar w)} e^{ip_3\cdot X(1)} e^{ip_4\cdot X(\infty)}\rangle  \nn
& =&  \intop d^{2}w\; |w|^{-4-\alpha' t/2} |1-w|^{-4-\alpha' s/2}  
\label{eq:A4}
\eea
In the Regge limit,  defined by   $|s|\rightarrow \infty$ along the positive imaginary axis, the second factor can be approximated by  $ |1-w|^{-4-\alpha' s/2} \sim e^{ \alpha' s(w+ \bar w)  /4}  $, 
leading  to  a cutoff,  $|w|=O(1/s)$.  
 This is equivalent to keeping the first non-trivial term,
\be
e^{\textstyle ip_1\cdot X(0)} e^{\textstyle ip_2\cdot X(w,\bar w)} \sim |w|^{\textstyle -4-\alpha' t/2} \; e^{
\textstyle i k\cdot X(0) + i p_2\cdot (w\dd + \bar w \cdot \bar \dd) X(0)}\; , \label{eq:OPE}
\ee
in the OPE. Regge behavior  then follows, 
\be
\int d^2 w|w|^{-4-\alpha' t/2}e^{ \alpha' s(w+ \bar w)  /4} = \Pi(\alpha(t) ) \;  s^{\alpha(t)}   \;,\label{eq:4pointRegge}
\ee
where 
\be
\Pi(\alpha( t) ) = 2\pi \frac{\Gamma(-\alpha(t)/2)}{\Gamma(1+\alpha(t)/2)}  e^{-i\pi \alpha (t)/2} \sim \frac{e^{-i\pi \alpha(t)} +1}{\sin \pi \alpha(t)}\ , \label{eq:evensignature}
\ee
and $\alpha(t) = 2 + \alpha' t/2$. 

We note that the terms $\dd X_M(0)$ and  $\bar \dd X_N(0)$ in (\ref{eq:OPE}) single out the level-one oscillators,  $a^\dagger_{1,M}$ and $\tilde a^\dagger_{1,N}$
that contribute to states on the leading Regge trajectory. We also note that, with (\ref{eq:evensignature}), Eq. (\ref{eq:4pointRegge}) can be expressed as   (\ref{eq:10dflatspacePomeron}), appropriate for an $C=+1$ exchange.

To generalize this to the $C= -1$ sector we must go beyond elastic  tachyon  amplitude. Due to symmetry  under crossing, $s \leftrightarrow u$, only $C = +1$ states, which are even under worldsheet parity, couple in the
t-channel. The odd sector couples first in the four point amplitude involving two external $C=-1$ states, one initial and one final in the t-channel.  This is due to world-sheet parity conservation.  It is also useful to consider the general case of scattering  involving an arbitrary number of external particles  of different types. The amplitude can be calculated at the tree-level as an integral
\begin{equation}
A_{n}=\intop d^{2}w_{2}d^{2}w_{3}\cdots d^{2}w_{n-2}<V_{1}V_{2}\cdots V_{N}>.\label{eq:}
\end{equation}
where  $V_{i}$ are the corresponding vertex operators, including that for $C=-1$ states. (Here, we have also made use of M{\"o}bius invariance to fix three points, i.e., $w_1$, $w_{n-1}$ and $w_n$.) In the Regge limit, one factorizes this  into left- and right-moving, denoted by  ${\cal W}_R$ and ${\cal W}_L$ for   sets of $l_R$ right-moving and $l_L$ left-moving vertex operators together with their associated $l_R-2$ and $l_L-2$ world-sheet integrations.
\be
 A_{n} = \int d^2w \left\langle {\cal W}_R w^{L_0 - 2}
   \bar w^{\bar L_0- 2} {\cal W}_L \right\rangle\ .
   \ee
Integrating over $\int d^2w$ leads to level matching
conditions $L_0=\bar L_0$ and the string propagator $(L_0 + \bar L_0 -2)^{-1}$.    It was also demonstrated in \cite{Brower:2006ea} that the leading Regge behavior  corresponds to satisfying the J-plane constraint,
\be
L_0(j,t)=\bar L_0(j,t)  =j/2    - \alpha' t/4 = 1, \quad {\rm or} \quad j=2+\alpha' t/2  \;. \label{eq:onshell}
  \ee
which interpolates through the physical states on the leading trajectory. Taken together these constraints can be represented 
by an inverse  Mellin transform, 
\be
A_{n}   \simeq  \int \frac{d j }{2\pi i}  \frac {{\cal F}(j) \; \delta_{L_0,\tilde L_0}} {L_0(j,t)+\tilde L_0(j,t) -2} \; .
\ee
The residue is evaluated by the insertion of Reggeon vertex operator
and the integral performed by contour integration over the pole at $j=2+\alpha' t/2$.  For example in the Pomeron case the   result~\cite{Brower:2006ea} was $A_{n} \sim \Pi(\alpha( t) ) \left\langle {\cal W}_{R}\, {\cal        V}_P^- \right\rangle \left\langle {\cal V}_P^+ \, {\cal W}_{L}    \right\rangle  $,  or more explicitly by a large relative boost, $exp[ -y M_{+-}]$, for  the states ${\cal W}_L$    and ${\cal W}_R$ back to their approximate rest frames (denoted by a    subscript 0),
\be
 A_{n} \sim \Pi(\alpha( t) ) (s/s_0)^{2 + \alpha' t/2} \left\langle
   {\cal W}_{R0}\, {\cal V}_P^- \right\rangle \left\langle {\cal V}_P^+ \,
   {\cal W}_{L0} \right\rangle \; . 
\label{pomgen2} 
\ee
where the large boost parameter (or rapidity gap) is $y = \log(s/s_0)$.  The result~(\ref{pomgen2}) has a simple interpretation as a Pomeron propagator, $ (s/s_0)^{2 + \alpha' t/2}$, times the couplings of the Pomeron to the two sets of vertex operators. To clarify this consider again the simple case of the 4-point function at $t=0$. In this case, one needs only to evaluate the following 3-point function,
\be
\langle e^{ip_1\cdot X(1)} e^{ip_2\cdot X(2)}e^{\mp i k\cdot X(w,\bar w)} (h_{MN} \dd X^{M}\bar \dd X^{N}/\alpha') \rangle\; ,
\ee
where $k=p_1+p_2$ and $k\cdot (p_1-p_2)=0$. Note that this is analogous to a graviton-tachyon-tachyon coupling vertex.   In CM, and using  LC coordinates, $p_{1,2}$ have large $``+"$ components, $O(\sqrt s)$, if they are R-moving, and  large $``-"$ components, if they are L-moving. The corresponding small components are $O(1/\sqrt s)$,  and components orthogonal to $\pm$ are $O(1)$. From the  commutation relations, $[a_1^{\pm} , {a^{\mp}_1}^\dagger] = \pm 1$,  the dominant contribution at large $s$ comes from $h_{-,-}$  (for R-moving) and $h_{+,+}$ (for L-moving), leading to the dominant LC components, (\ref{eq:Pomeronvertexop}).  The analysis  above holds also for $t\neq 0$ and can be carried out  for general n-point amplitudes,  leading to (\ref{pomgen2}).

Returning to the general case insert the operator,
\be
{\cal V}(T) = (T_{MN} \dd X^{M}\bar \dd X^{N}/\alpha')^{1+\alpha' t/4} e^{\mp i k\cdot X}
\ee 
and factorize the tensor and take the high energy limit. The leading term
on-shell at $t=0$ defines  3 vertex operators:
 \bea
 \langle {\cal  W}_R \; e^{ikX}    h_{MN}\partial X^{M} \overline{\partial}X^{N}\rangle &\simeq &  \langle {\cal  W}_R \; e^{ikX}    h_{--}  \partial X^{-}\overline{\partial}X^{-}   \rangle  = O(s)\; ,\nn
 \langle {\cal  W}_R \; e^{ikX}    B_{MN}\partial X^{M}\overline{\partial}X^{N}\rangle &\simeq &  \langle  {\cal  W}_R\; e^{ikX}    B_{-\perp}( \partial X^{-}\overline{\partial}X^{\perp}-\partial X^{\perp}\overline{\partial}X^{-})\rangle  = O(\sqrt{s})\;,\nn
  \langle {\cal  W}_R \;  e^{ikX}  \eta_{MN}\partial X^{M}\overline{\partial}X^{N}\rangle &\simeq & O(1)\; .
  \eea
For   $t\ne 0$ the leading term picks up a common
factor of $(\partial X^- \bar \partial X^-)^{\frac{\alpha't}{4}} $ on the  left,
\bea
[(h_{MN}&+&B_{MN})\partial X^{M}\overline{\partial}X^{N}]^{1+\frac{\alpha't}{4}}\\
&\simeq&\left[  h_{--}  (\partial X^{-}\overline{\partial}X^{-})+B_{-\perp} (\partial X^{-}\overline{\partial}X^{\bot}-\partial X^{\perp}\overline{\partial}X^{-})\right] (\partial X^{-}\overline{\partial}X^{-})^{\frac{\alpha't}{4}} \; . \nonumber
\eea
For ${\cal W}_L$, we need to replace  $X^{-}$ by $X^+$,  with a common factor $(\partial X^+ \bar \partial X^+)^{\frac{\alpha't}{4}} $ on the right.  When they are combined, the symmetric combination, i.e., the first term above,  leads to the Pomeron vertex operator. The anti-symmetric combination, the second term, contributing to $1/ s$ down relative to the first term, corresponds to an Odderon exchange. 

In analogy with ${\cal V}_P^{\pm}$, we can now define an Odderon vertex operator
\be
{\cal V}_O^{\pm} =(2\epsilon_{\pm,\perp}\dd X^{\pm}\bar \dd X^{\perp}/\alpha') (2\dd X^{\pm}\bar \dd X^{\pm}/\alpha')^{\alpha' t/4} e^{\mp i k\cdot X}
\ee
which characterizes the exchange of a leading Odderon.  Here $\epsilon_{\pm,\perp}=-\epsilon_{\perp,\pm}$.
Just as for the Pomeron, this is an on-shell vertex operator, satisfying the on-shell conditions
\be
L_0 {\cal V}^{\pm}_O= \bar L_0{\cal V}^{\pm}_O = {\cal V}^{\pm}_O
\ee
In particular, this  leading Regge trajectory interpolates those states created by level-one oscillators, $a_{1,M}^\dagger$ and $\tilde a_{1,M}^\dagger$, with   $ j = 2n+1$, $t_n =  4 n \; {\alpha'}^{-1}$,  $ n=0,1,\cdots$, 
and can be characterized by an  Odderon trajectory
 \be
 \alpha^{(-)}(t) = 1 + \alpha' t/2\;, \quad {\rm or } \quad  L_0(j,t)=\bar L_0(j,t)  =(j+1)/2    - \alpha' t/4 = 1
 \ee
Again we may boost the states back to their rest frame,
\be
 A_{n} \sim  \Pi_-(\alpha^{(-)}(t)) (s/s_0)^{ \alpha^{(-)}(t) } \left\langle 
   {\cal W}_{R0}\, {\cal V}_O^- \right\rangle \left\langle {\cal V}_O^+ \,
   {\cal W}_{L0} \right\rangle \; . 
\label{eq:oddgen} 
\ee
to explicitly obtain the Odderon Regge amplitude.  Here, $\Pi_{-}(\alpha)= \Pi(\alpha + 1)$, and (\ref{eq:oddgen}) is of the form (\ref{eq:10dflatspaceOdderon}), appropriate for an crossing-odd exchange.

Let us next briefly comment on the the couplings of the Odderon, i.e., 
$   \langle {\cal W}_{R0}\, {\cal V}_O^- \rangle $ and  $\left\langle {\cal V}_O^+ \,
   {\cal W}_{L0} \right\rangle$.
Note that ${\cal V}_P$ and ${\cal V}_O$ have opposite symmetry under $\dd \leftrightarrow \bar\dd$, even and odd for   ${\cal V}_P$ and ${\cal V}_O$ respectively.  For oriented closed strings, the theory should be invariant under $w \leftrightarrow \bar w$. It follows that non-vanishing coupling $   \langle {\cal W}_{R}\, {\cal V}_O^- \rangle $ and $   \langle {\cal W}_{L}\, {\cal V}_O^+ \rangle $  would require ${\cal W}_{R,L}$ to be odd under $w \leftrightarrow \bar w$. For example, tachyon and graviton vertex operators   $V_t$ and $V_g$ are   even under $w\leftrightarrow  \bar w$. It follows that couplings involving n-tachyons and an Odderon vanish.  To be precise, $  \langle {\cal W}_{R}\, {\cal V}_O^- \rangle $ and $  \langle {\cal W}_{L}\, {\cal V}_O^+ \rangle$ are non-zero only if ${\cal W}_{R,L}$ are odd under $w\leftrightarrow  \bar w$. Conversely, $  \langle {\cal W}_{R}\, {\cal V}_P^- \rangle $ and $  \langle {\cal W}_{L}\, {\cal V}_P^+ \rangle$ are non-zero only if ${\cal W}_{R,L}$ are even under $w\leftrightarrow  \bar w$.    It is worth mentioning that our discussion for the Pomeron and Odderon vertex operators also apply to couplings with branes, which can be used to model ``mesons''. It should also apply in the case of  baryons. One can also generalize to the Fermionic sector, as discussed briefly in \cite{Brower:2006ea}.

\subsection{Pomeron and Odderon in AdS}

We next extend the flat-space  results  to the case of an $AdS$ background. We will  focus on 
$AdS_{5}$ without confinement deformation. 
In the limit of large $\lambda$, ($\lambda \equiv {R^{4}}/{\alpha'^{2}}$),  the world sheet path integral will be Gaussian. As emphasized in \cite{Brower:2006ea} for the Pomeron, we need to generalize the  vertex operator to include a  string wave function in $AdS$. In enforcing the on-shell condition, we must diagonalize $L_0$ and $\tilde L_0$ by going  to a basis of definite spin. As we demonstrate below, making use of  the analysis in Sec. \ref{sec:diffusion}, a complete diagonalization can be carried out.

For the $C=+1$ sector,  let us begin by considering   the vertex operator in a definite spin basis to have the form
\begin{equation}
V_P( j,\pm)=(\partial X^{\pm}\overline{\partial}X^{\pm})^{\frac{j}{2}}e^{\mp ik\cdot X}\phi_{\pm j}(r).\label{eq:PomeronvertexAdS}
\end{equation}
From the physical state conditions, $L_{0}V_P(j,\pm)= \tilde L_{0}V_P(j,\pm)=V_P(j,\pm)$, one has, up to $O(1/\sqrt \lambda)$ corrections, $ [\frac{j-2}{2}-\frac{\alpha'}{4}\triangle_{j}]e^{\mp ik\cdot X}\phi_{\pm j}(Y)= 0$.  The on-shell condition can again be expressed in terms of a Mellin transform,
\be
\int \frac{dj}{2\pi i} \frac {\delta_{L_0,\tilde L_0}} { L_0 (j)+ \bar L_0(j) -2}\;.
\ee
It is  convenient to  first perform a similarity transformation, $(r/R)^{(j-2)}[ L_0 (j)+ \bar L_0(j) -2]^{-1} (r/R)^{-(j-2)}$, leading to  the $J$-plane Pomeron kernel is 
\be
G^{(+)}(j) = \frac {1}{ j-2 -( {R^2}/{2\sqrt \lambda}) \Delta_{2}}\;,
\ee
and   we have replaced  $\alpha'$ by $R^2/\sqrt\lambda$.  Note that this is precisely the Pomeron propagator introduced earlier,   (\ref{eq:pomeronpropagator}), by a diffusion consideration.

As shown in Sec. (\ref{sec:diffusion}), $\Delta_2$ can  be diagonalized using the $u$-basis, with the introduction of another continuous quantum number $\nu$.  In this basis, 
$ \phi_{+j} \sim e^{(j-2)u} K_{2i\nu} (|t|^{1/2} e^{-u}) \; , $ 
${\cal D}=2/\sqrt \lambda$, and the Pomeron vertex operator takes on the form
 \be
 {\cal V}_P(j,\nu,k,\pm) \sim (\partial X^{\pm}\overline{\partial}X^{\pm})^{\frac{j}{2}}e^{\mp ik\cdot X}e^{(j-2)u} K_{\pm 2 i \nu} (|t|^{1/2} e^{-u}) \; .
\ee
(More discussion for these  solutions  can be found in \cite{Brower:2006ea}.)  Following next  the same steps done for the flat space, e.g., boosting back to the respective rest  frames of $L$- and $R$-moving particles, we arrive at, for $C=+1$, 
\bea
{\cal T}^{(+)} 
   &\sim& \int \frac{dj }{2\pi i} \int \frac{d\nu \nu\sinh{ 2\pi \nu} }{ \pi}  \frac{ \Pi(j)\; s^{j} }{ j- j_0^{(+)} + {\cal D} \nu^2}\left\langle
   {\cal W}_{R0}\, {\cal V}_P(j,\nu,k,-) \right\rangle \left\langle {\cal V}_P(j,\nu,k,+) 
   {\cal W}_{L0} \right\rangle ,   \nn
\eea
where $\Pi(j)$ leads to a signature factor, appropriate for ${\cal T}^{(+)} $ being even in $s$.
Closing the contour at $\nu =i \lambda^{1/4} \sqrt{ (j- j_0^{(+)})/2}   $ directly leads to the desired final result obtained in \cite{Brower:2006ea}.

We are now in the position to generalize to  the case of  Odderon by following the same steps.   Going to the $J$-plane, we introduce  a vertex operator
\be
{\cal V}_O(j,\pm)=(\partial X^{\pm}\overline{\partial}X^{\perp}-\partial X^{\perp}\overline{\partial}X^{\pm})(\partial X^{\pm}\overline{\partial}X^{\pm})^{\frac{j-1}{2}}e^{\mp ik\cdot X}\phi_{\pm j\perp }(r).
\label{eq:OdderonvertexJ}
\ee
 The physical state condition, $L_{0}V^\pm_O(j)=\tilde L_{0}V^\pm_O(j)=V^\pm_O(j)$,  will then give us
\be
[\frac{j+1}{2}-\frac{\alpha'}{4}\Delta_{O,j}] \phi_{\pm j\bot}(r)=\phi_{\pm j\bot}(r). \label{eq:phys-odd-ads}
\ee
We can again perform a similarity transformation, $\phi_{\pm j\perp} = (r/R)^{(j-1)} \phi_{\pm,\perp}$, and arrive  at a $J$-plane Odderon propagator, 
\be
G^{(-)}(j) = \frac {1}{ j-1 - ({R^2}/{2\sqrt \lambda}) \Delta_{O,1}}\;,
\ee
where $ \Delta_{O,j}= (r/R)^{-(j-1)}  ( \Delta_{O,1} ) (r/R)^{(j-1)}$.
To determine the diffusion operator for Odderon, $\Delta_{O,1}$, we can match the EOM at $j=1$, appropriate in the infinite $\lambda$ limit.  

The EOM in this case involves a Maxwell operator,
and in general will also involve an AdS mass.  Both the Maxwell operator and the AdS mass squared, $m_{AdS}^{2}$, 
can be fixed by  a proper identification of the
$B_{\mu\nu}$ state of string theory at $j=1$ using the $AdS/CFT$ dictionary \cite{Brower:2000rp}.  
In \cite{Kim:1985ez,Brower:2000rp} it was found that a 2-form fields in AdS space
has two solutions, corresponding to  $m_{AdS}^{2}=16$  and  $m_{AdS}^{2}=0$.

  As done for the Pomeron, to $O(1/{\sqrt\lambda})$,  after writing  out the Maxwell
operator, we have, for $z\neq z'$, and for $j\simeq 1$, $G^{(-)}(j)$ satisfies the $C=-1$ part of Eq. (\ref{eq:DE2}).
Again, the wave function, $\phi_{\pm\perp}$, satisfies the same DE.  Changing  variable to
$u=-\ln (z/z_0)$, we get 
\be
[j-1-\frac{\alpha' t }{2}e^{-2u}  -\frac{1}{2\sqrt{\lambda}}(\partial_{u}^{2}-m_{AdS}^{2})]\phi_{\pm \bot}(u)=0.
\ee
This equation can again be diagonalized by $\nu$, with
 \be
 \phi_{+j\perp} \sim e^{(j-1)u} K_{2i\nu} (|t|^{1/2} e^{-u}) \; , \quad {\rm and}  \quad G^{(-)}(j) = \frac {1}{ j- j_0^{(-)} +  {\cal D} \nu^2 }\;.
 \ee
where $j_0^{(-)}= 1 - m^2_i/2\sqrt\lambda$, Eq. (\ref{eq:dualconformalOdderon}). In this basis, the Odderon vertex operator takes on the form
 \be
 {\cal V}_O(j,\nu,k,\pm) \sim (\partial X^{\pm}\overline{\partial}X^{\perp}-\partial X^{\perp}\overline{\partial}X^{\pm})(\partial X^{\pm}\overline{\partial}X^{\pm})^{\frac{j-1}{2}}e^{\mp ik\cdot X}e^{(j-1)u} K_{\pm 2 i \nu} (|t|^{1/2} e^{-u})
\ee
Following the same steps done for $C=+1$, we arrive at
\bea
{\cal T}^{(-)}  &\sim& \int \frac{dj }{2\pi i} \int \frac{d\nu \nu\sinh {2 \pi \nu} }{\pi}  \frac{ \Pi_-(j)\; s^{j}}{ j- j_0^{(-)} + {\cal D} \nu^2}\left\langle
   {\cal W}_{R0}\, {\cal V}_O(j,\nu,k,-) \right\rangle \left\langle {\cal V}_O(j,\nu,k,+) \,
   {\cal W}_{L0} \right\rangle    \nn 
\eea
where we again have  a signature factor appropriate for amplitude being odd in s. Closing the contour at $\nu =i \lambda^{1/4} \sqrt{  (j- j_0^{(+)})/2}   $ directly leads to the desired final result.

Recall that there are  two solutions for the conformal Odderon. For $m^2_{AdS}=16$,
\be
j^{(-)}_{0,(1)}=1- 8/\sqrt{\lambda} + O(1/\lambda)\;,   \label{eq:dualconformalOdderon1}
\ee
and, for $m^2_{AdS}=0$, 
\be
j^{(-)} _{0,(2)}=1+ O(1/\lambda)\;, \label{eq:dualconformalOdderon2}
\ee
as summarized in Table \ref{tab:intercepts}.

\subsection{Conformal Dimensions, and BFKL/DGLAP Connection}\label{sec:general}

We end this section with a brief discussion on the quantity $\Delta^{(\pm)}(j)$, defined in Eq. (\ref{eq:Delta4}),  which sets the dimension $\Delta$ as a function of Lorentz spin $J_{Lorentz}$ for the BFKL/DGLAP operators for $C=\pm 1$ sector respectively.  This connection was first emphasized in Ref. \cite{Brower:2006ea}. The analytic continuation from DGLAP to
BFKL operators has been discussed at weak coupling for some time
\cite{Jaroszewicz:1982gr,Lipatov:1996ts,Kotikov:2000pm,Kotikov:2002ab}.  Recently, it was conjectured to be exact at weak coupling
in ${\cal N}=4$ Yang-Mills theory \cite{klv5}.  The
demonstration of
this relationship in all large-$\lambda$ conformal theories, and the
derivation of the formula for $\Delta^{(+)}(j)$, is given in section 3 of
\cite{Brower:2006ea}, where
the existence of the single function
$\Delta^{(+)}(j)$ with $j=j^{(+)}_0$ at $\Delta=2$ (the BFKL exponent) and $j=2$ at
$\Delta = 4$ (for the energy-momentum tensor, the first DGLAP
operator) was demonstrated.

The AdS/CFT dictionary relates string states to local operators in the gauge theory. This can be summarized by the statement that gauge theory correlators can be found, in the strong coupling, by behavior of bulk (gravity) fields, $\phi_i(x,z)$,  as they approach the $AdS_5$ boundary, $z\rightarrow 0$,
\be
\langle e^{\int d^4x \phi_i(x) {\cal O}_i(x)}\rangle_{CFT} = {\cal Z}_{string} \left[ \phi_i(x,z) |_{z\sim 0}  \rightarrow   \phi_i(x) \right] \;, 
\ee
where $z>0$ extends  into the $AdS$ bulk and the left hand side is the generating function of correlation functions in the gauge theory for the operator ${\cal O}_i$. For instance, in the case where $\phi(x,z)$ is the dilaton field in the bulk, it is well-known that the corresponding gauge theory operator ${\cal O}$ is $ Tr F^2$.

Let us next consider the diagonalized Pomeron and Odderon vertex operators. For simplicity, we consider  the $t=0$ limit, where they   become:
\bea
 {\cal V}_P(j,\nu,0,\pm) &\sim& (\partial X^{\pm}\overline{\partial}X^{\pm})^{\frac{j}{2}}e^{(j-2 \pm2 i\nu )u} \\
  {\cal V}_O(j,\nu,0,\pm) &\sim&  (\partial X^{\pm}\overline{\partial}X^{\perp}-\partial X^{\perp}\overline{\partial}X^{\pm})(\partial X^{\pm}\overline{\partial}X^{\pm})^{\frac{j-1}{2}}e^{(j-1\pm 2i\nu)u} 
   \eea
These vertex operators for integer $j\geq 2$ and $j\geq 1$ respectively  correspond to the lightest string states of given spin, and so are dual to the lowest dimension operators of those spins. Therefore, in the large $\lambda$ limit, the physical state conditions, $L_0=\tilde L_0=1$, would determine the dimensions of the leading twist operators.   Let us examine the Pomeron and the Odderon vertex operators from this perspective. 

For $C=+1$,  ${\cal V}_P$ at the AdS boundary  will couple to a gauge invariant operator ${\cal O}^{(+)}$, i.e., 
\be
\int d^4x\; {\cal O}^{(+)} \; {\cal V}_P
\ee
where ${\cal O}^{(+)}$ is an operator of  dimension $\Delta^{(+)}$ and Lorentz spin $J_L$.  By examining the number of Lorentz indices, we have  $J_L=j$.  At $ j=2$, ${\cal V}_P$ couples to the energy-momentum tensor, whose gauge part is $tr(F_{+\mu} F_+^\mu)$.  For general $j$, as a leading twist operator, its  gauge part  is  $tr(F_{+\mu} D_+^{J_L-2} F_+^\mu)$, with the twist given by
\be
\tau^{(+)} = \Delta^{(+)} - J_L
\ee
At $j=2$,  $\Delta^{(+)}(2) = 4$, corresponding to  twist two. Due to energy-momentum conservation, anomalous dimension vanishes there,  as expected.

\begin{figure}[h]
\begin{center}
\includegraphics[width = 0.8\textwidth]{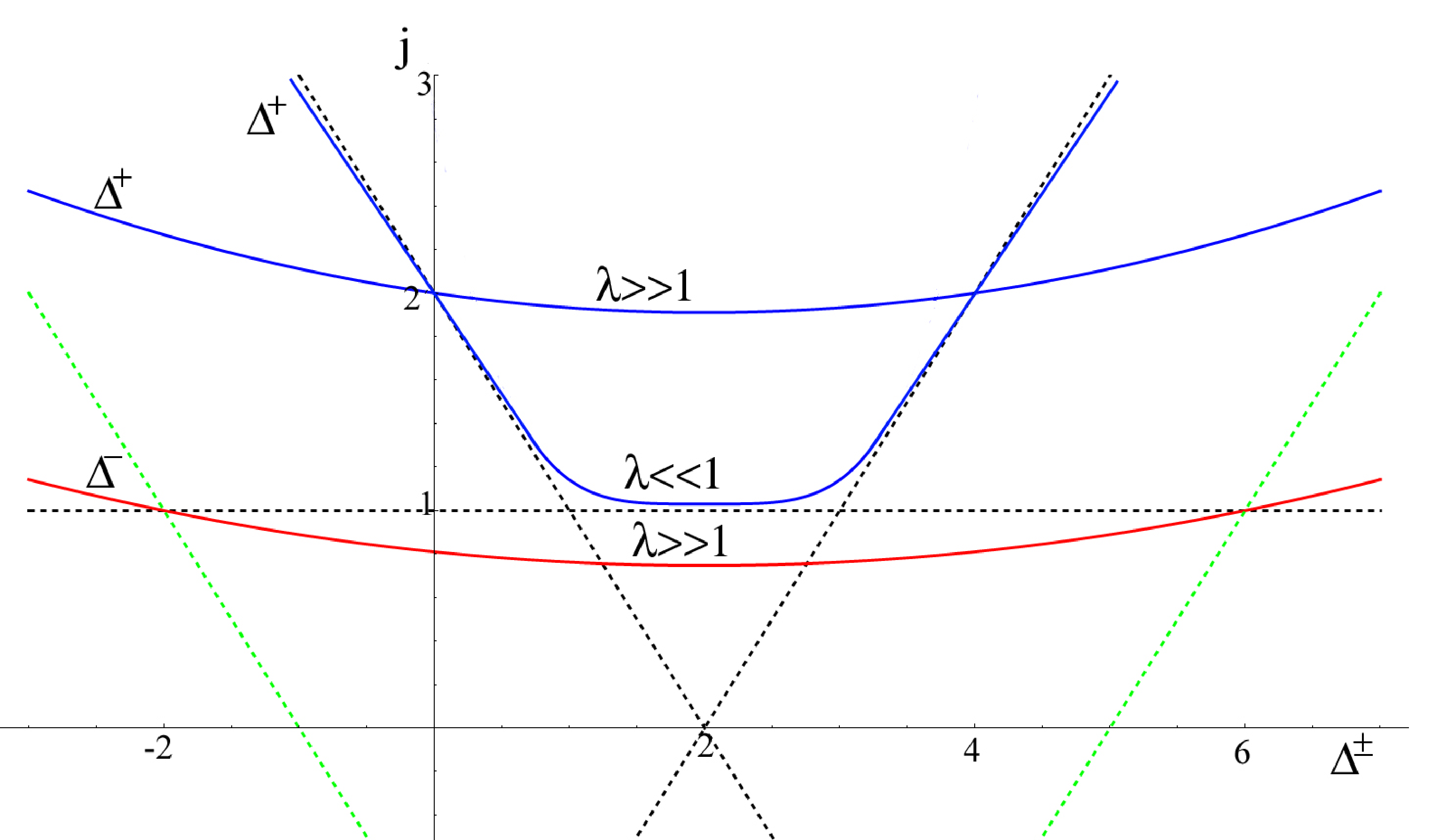}
\caption{Schematic form of the $\Delta^{(\pm)}-j$ curves.   For $C=+1$, we show in blue  the curves ($\Delta^{(+)}$) for both  the limit  $\lambda \gg  1$ and $\lambda\ll 1$.  (Fig. 2 of Ref. \cite{Brower:2006ea}.)   The curve for $C=-1$ ($\Delta^{(-)}$) at $\lambda \gg 1$ is shown in red,  for the branch $m^2_{AdS} =16$. The dashed lines show the $\lambda =0$ DGLAP branch (slope 1), BFKL branch (slope 0), and inverted DGLAP branch (slope $-1$) for both $C=\pm 1$.  Note that the $C= +1$ curves pass through the points (4,2) and
(0,2) where the anomalous dimension must vanish.    For $C=-1$,  the curve pass through the points (6,1) and (-2,1).   The inversion symmetry $\Delta \to 4-\Delta$ is shown explicitly.  Note also that, Lorentz spin and $SO(3)$ spin are related by  $J_L=j$ and  $J_L=j+1$ for $C=+1$ and $C=-1$ respectively.}
\label{fig:BFKLDGLAP}
\end{center}
\end{figure}

To determine $\Delta^{(+)}(j)$, let us first examine the scale transformation for ${\cal V}_P$. Under scale transformation, if  $\delta u \sim \epsilon$, then $\delta X\sim - \epsilon X$. If $\phi_{j+}\sim e^{\zeta u}$, $\zeta=j-2+2 i\nu$, it follows ${\cal V}^{+}_P$ scales with a dimension $2 -2 i\nu$.
Therefore~\footnote{As noted earlier, $\nu$ used here is half of that used in Ref. \cite{Brower:2006ea}.}
\be
\Delta^{(+)} = 4- (  2 - 2 i\nu)=2+ 2i\nu
\ee
On the other hand,  from 
\be
j= j_0^{(+)} - {\cal D} \nu^2 
\ee
one has
\be
\Delta^{(+)} (j)= 2 +\sqrt{ 2} \; \lambda^{1/4} \sqrt{ j-j_0^{(+)}}
\ee
or 
\be
j=  j_0^{(+)} + (1/2\sqrt\lambda )(\Delta^{(+)}-2)^2 
\ee
 as depicted in Fig. \ref{fig:BFKLDGLAP}.
As noted earlier,   $\Delta = 4$ for $J_L=j=2$, due to energy-momentum conservation.  Note also  that the the curve is symmetric  in $\Delta\leftrightarrow 4-\Delta$, reflecting inversion symmetry in $AdS_5$.  The intercept for the conformal Pomeron can be read off as the minimum $j$-value at $\Delta =2$. (See Table \ref{tab:intercepts}.)

We can repeat the same step for $C=-1$. One again arrives at
\be
\Delta^{(-)}= 2+ 2 i\nu
\ee
The corresponding $\Delta-j$ curve  is
\be
j=j_0^{(-)} +  (1/2\sqrt\lambda ) (\Delta^{(-)}-2)^2 
\ee
or
\be
\Delta^{(-)} (j)= 2 +\sqrt{ 2}\; \lambda^{1/4}\; \sqrt {j-j_0^{(-)}}
\ee
Note that  the curve is again symmetric about  $\Delta^{(-)}=2$.  The Odderon intercept can again be read off as the minimum $J$-value at $\Delta^{(-)} =2$. We also note, at $j=1$, ${\cal V}_O$ has Lorentz spin 2, $J_L=2$, thus $j=J_L-1$. 

There are now two branches. Consider first the branch where $m^2_{AdS} = 16$.  At  $j=1$, one has $\Delta^{(-)}=6$. With  Lorentz spin 2, it follows that this corresponds to a  twist-4 operator, whose gauge component can be identified with    $tr(F_{+\perp} F^2)$.  Due to charge conservation, anomalous dimension again vanishes there. For  general $J_L$, ($J_L=j+1$), $ {\cal V}_O$  couples to    $tr(F_{+\perp}D_+^{j-1} F^2)=   tr(F_{+\perp}D_+^{J_L-2} F^2)$.   For $\lambda\neq  \infty$, operator mixing naturally  leads to the curve shown in Fig. \ref{fig:BFKLDGLAP}, with $j_0^{(-)}<1$, consistent with that found in the weak coupling.    For the branch $m^2_{AdS}=0$, we have $\Delta^{(-)}=2$ at $j=1$, and it would be interesting to identify the corresponding gauge invariant operators  $ {\cal V}_O$ for various integral $j$ values.

\newpage

\section{Effects of Confinement in the Hardwall Model}
\label{sec:spectrum}

So far, we have carried out our discussion mostly  in the conformally invariant limit, or in the kinematic regimes where confinement played no significant role. More generally, the Maldacena duality conjecture and its further extensions  assert that there is an exact equivalence between large $N_c$  conformal field theories in d-dimensions and string theory in $AdS^{d+1}\times M$.  A dual gravity description for $SU(N)$ quarkless $QCD_4$ was first suggested by  Witten \cite{Witten:1998zw} by breaking explicitly the conformal (and SUSY) symmetries. A systematic strong coupling glueball calculation within the Witten scheme \cite{Brower:2000rp} has been carried out  in the supergravity limit, i.e., infinite $\lambda$ limit, and the resulting spectrum is in qualitative agreement  with ground state levels from lattice $QCD_4$~\footnote{
Due to mixing with ordinary mesons, experimental identification of
glueball states has been challenging.  The best evidence for their
existence has been through lattice gauge
theory~\cite{Morningstar:1999rf}.  For first attempts at calculating glueball masses
using AdS/CFT, following work of \cite{Witten:1998zw},
see \cite{Csaki:1998qr, deMelloKoch:1998qs,Hashimoto:1998if,Aharony:1999ti}. The
relevant tensor glueball state was first studied
in~\cite{Brower:1999nj,Brower:2000rp,Constable:1999gb}.}.  In particular, it reproduces the important feature for low mass glueballs: 
\be
m^2(0^{++}) < m^2(0^{-+})\sim m^2(2^{++})\;.
\ee
(See Fig. \ref{fig:comparison} below.) It has also been noted in \cite{Brower:2000rp} that  the feature: $ m^2(0^{++}) < m^2(0^{-+})\sim m^2(2^{++}) < m^2(1^{+-})< m^2(1^{--})$, is  consistent with  the  pattern derived from a  ``constituent gluon'' or bag models for glueballs. 
\begin{figure}[h]
\includegraphics[height = 0.5\textwidth,width = 0.4\textwidth]{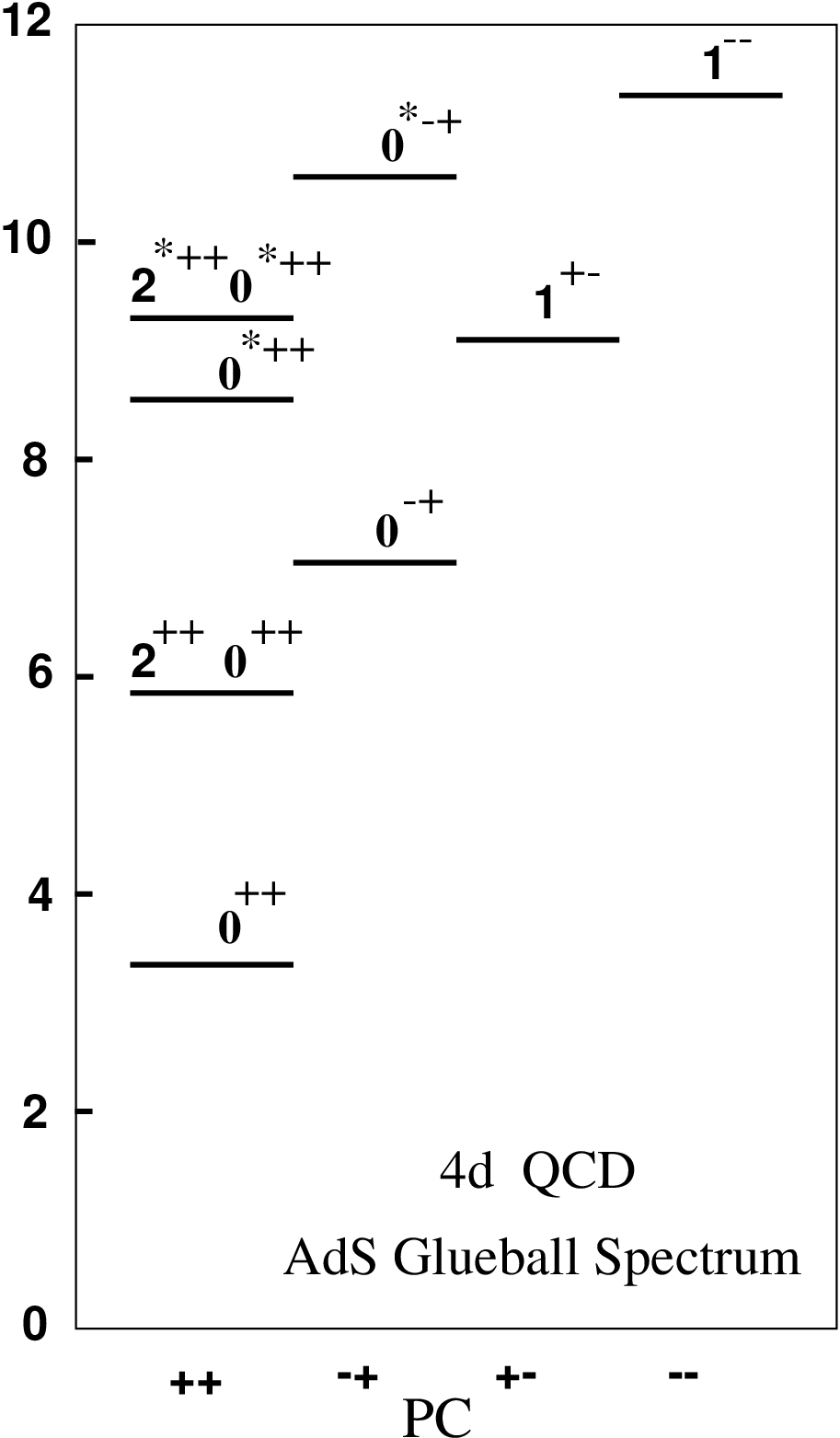}
\qquad
\includegraphics[height=0.52\textwidth,width=0.5\textwidth]{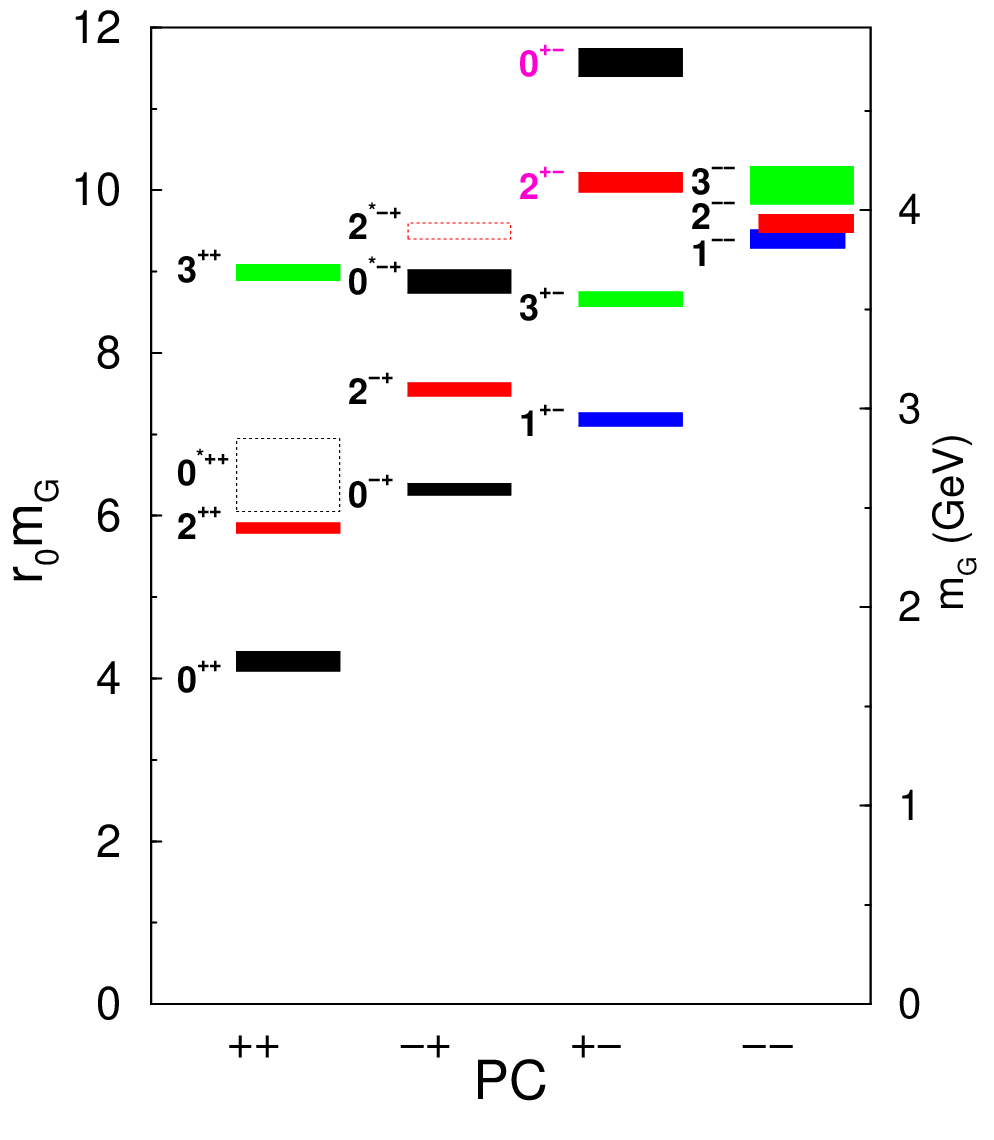}
\caption{The AdS glueball spectrum for $QCD_4$ in strong coupling   (left) compared with the lattice spectrum~\cite{Morningstar:1999rf}   for pure SU(3) QCD (right). The AdS cut-off scale is adjusted to set   the lowest $2^{++}$ tensor state to the lattice results in units of   the hadronic scale $1/r_0 = 410$ Mev.  This is Fig. 2 of Ref.   \cite{Brower:2000rp}, which we reproduce it here.} \label{fig:comparison}
\end{figure}

The Witten proposal, however, starts with 11-dimensional M theory on $AdS^7\times S^4$. One of the dimensions, $x_{11}$, is taken compact, reducing the theory to type-IIA string theory. The 5-d Yang-Mills CFT is next dimensionally reduced to $QCD_4$ by raising the ``temperature'', $\beta^{-1}$, in a direction $x_5=\tau$. The new metric is an $AdS^7$ black hole with $x_{11}$ compact~\footnote{A similar calculation has also been carried out for   $QCD_3$, where one starts with $AdS_5$, dimensionally reduced by raising the   temperature to $\beta^{-1}$, leading to an $AdS_5$ black hole background   \cite{Brower:1999nj}.}  In spite of the qualitative success of the low mass glueball spectrum in this background, it has the weakness of a model QCD having incorrect conformal limit in the ultraviolet. The correct metric for 4-d Yang Mills theory is an $AdS_5$ background with mild deviations to account for the logarithmic running of asymptotic freedom.

Other efforts in calculating glueball masses using more realistic background, e.g.,  the Klebanov-Strassler background \cite{KS},   include \cite{Caceres:2000qe,PandoZayas:2003yb,Amador:2004pz,Caceres:2005yx,Dymarsky:2008wd}.  These backgrounds are more difficult to handle so in the current effort for simplicity, we shall work the $AdS_5$ metric, (\ref{AdSWmetric}), with a hard cutoff at $r=r_0$, i.e., a hardwall model.  This metric does not satisfy the supergravity equations, but experience has shown \cite{Polchinski:2001tt,DIS,rhouniv} that it captures both the phenomenology encoded in the metrics of consistent four-dimensional theories with confining dynamics \cite{nonestar,KS} and the qualitative properties of a near conformal theory in the UV.  In particular, phenomena for which the details of the metric in the confining region are not important --- potentially universal features of gauge theory --- are often visible in this model.  One can identify infrared-insensitive quantities and general features of the hadronic spectrum, hadronic couplings, etc, including, as we will see, aspects of Regge trajectories and of the Pomeron, Odderon, etc.  Meanwhile, model-dependent aspects of these and other phenomena can also be recognized, through their sensitivity to small changes in the model. The prices you pay for this phenomenological IR cut-off is the freedom to modify the boundary condition at the hardwall as part of detailed implementation of the cut-off.

\subsection{Spin, Degeneracy  and EOM at $\lambda= \infty$}

We will first consider glueball spectra in the supergravity limit, ($\lambda \rightarrow \infty$), before addressing the issue of the associated Regge trajectories, ($\lambda$ large but finite.) Type-IIB string theory at low energy has a supergravity multiplet with several zero mass bosonic fields: a graviton $G_{\mu\nu}$, a dilaton $\phi$, an axion (or zero form RR field) $C_0$, and two K-R tensors, the NS-NS and R-R fields $B_{\mu\nu}$ and $C_{2,\mu\nu}$ respectively. With a hard cutoff for $AdS_5$, modes  for these fields become discrete. 

Our first task is to find out all quadratic fluctuations in the $AdS_5\times S^5$ background metric whose eigen-modes correspond to the discrete glueball spectra for $QCD_4$ at strong coupling for various $J^{PC}$ sectors. We are only interested in the excitations that lie on the superselection sector for $QCD_4$. Thus for example we can ignore all non-trivial harmonic in $S^5$ that carry a non-zero R charge. The result of these considerations, discussed in detail below, is  summarized in Table \ref{tab:classIIB}.

To count the number of independent fluctuations for a supergravity field, we imagine harmonic plane waves propagating in the $AdS$ radial direction, $r$, with Euclidean time, $x_4$. For example, the metric fluctuations in $AdS_5$
\be
G_{\mu\nu}= \bar G_{\mu\nu} + h_{\mu\nu}(x)
\ee
in the fixed background $\bar G_{\mu\nu}$ are taken to be of the form $h_{\mu\nu}(r,x_4)$. There is no dependence on the other spatial coordinates, $\vec x= (x_1,x_2,x_3)$. As we shall explain in more details in Appendix A, there are 5 independent transverse metric fluctuations, $h_{ij}$, (spin 2), 3 each for NS-NS  and R-R two-form tensors respectively, $B_{ij}$ and  $C_{2,ij}$, (spin 1), 1 for the dilaton $\phi$ (spin 0), and  1 for the Axion $C_0$ (spin 0).  

We can consider perturbations of the following forms:
\bea
h_{ij}(r,x_4) &= & a_{ij}\;  t(r) \; e^{ik_4 x_4} \; ,  \\
B_{ij}(r,x_4) &=&  b_{ij}\; q_1(r) \;e^{ik_4 x_4}    \; , \\
 C_{2,ij}(r,x_4) &=&  c_{ij}\; q_2(r) \; e^{ik_4 x_4}\; \\
 \phi(r,x_4) &=& S_1(r) \;e^{ik_4x_4}  \; , \\   
C(r,x_4) &=& S_2(r) \; e^{ik_4x_4}  \; ,
 \eea
where $k=(0,0,0,k_4)$, $a_{ij} $ a constant  traceless-symmetric $3\times 3$ tensor, and $b_{ij}$, $c_{ij}$ anti-symmetric, with $i,j=1,2,3$.  Let $m^2=-k^2=-k_4^2$. From linearized Einstein's equations for $h_{ij}$, scalar wave-equations for $\phi$ and $C_0$, and a set of Maxwell equations for $B$ and $C_2$, we arrive at
\bea
[ - {r} \frac{d}{dr} r \frac{d}{dr}   + 4 ] \; t(r) &=& \frac{m^2}{r^2} t(r) \;,  \label{eq:h1} \\  
   \left[ - {r} \frac{d}{dr} r \frac{d}{dr}   + m^2_{AdS} \right] q_i (r) &=& \frac{m^2}{r^2} q_i (r) \; ,\\
     -  \frac{d}{dr} r^5 \frac{d}{dr} \; S_i(r)& = &  {m^2}\; {r}\; S_i(r)\; , 
\eea
where $m^2_{AdS}$ are $AdS$ mass squared for the two-form.   Note that, for $h_{ij}$ and $S_i$, $m^2_{AdS}=0$. We can bring all three equations into  the ``scalar'' form by scaling
\bea
T(r) &=&  r^{-2} \; t (r) \; , \nn
Q_i(r) &=& r^{-2}\; q_i(r) \; .
\eea
One finds that, for $T$,  $Q_i$ and $S_i$, 
\bea
 -   \frac{d}{dr} r^5 \frac{d}{dr}T(r) & = &  {m^2}\;{r} \;T(r) \;,\nn
  \left[-  \frac{d}{dr} r^5 \frac{d}{dr}   + (m^2_{AdS}-4) r^3 \right] Q_i(r) &=&  {m^2}\;{r}\; Q_i(r) \; , \label{eq:scalarform} \\
      -  \frac{d}{dr} r^5 \frac{d}{dr} S_i(r) &=&  {m^2}\;{r}\; S_i(r)  \;. \nonumber
\eea
Note that equations for $T$ and $S_1$  and  $S_2$ are degenerate, and also for $Q_1$ and $Q_2$.
We will find it  convenient later  to work directly with $t$, $q_i$ and $s_i$, where $s_i= r^2 S_i$. One finds an equivalent set of equations
\bea
\left[ -  r \frac{d}{dr} r \frac{d}{dr} +4\right] t(r) & = &  (m^2/r^2) \; t(r) \;,\nn
  \left[-  r \frac{d}{dr} r\frac{d}{dr}  + m^2_{AdS} \right] q_i(r) &=&  (m^2/r^2) \; q_i(r) \; ,\nn
 \left[-  r \frac{d}{dr} r \frac{d}{dr} +4\right] s_i(r) &=&   (m^2/r^2)  \; s_i(r)  \;.   \label{eq:reducedform}
\eea

For definiteness, let us for now work with Eqs. (\ref{eq:scalarform}). Normalizability also requires wave functions to vanish at $r\rightarrow \infty$. (The case $m^2_{AdS}=0$ requires a special treatment~\footnote{ In that case, $Q=r^{-2} \log(r/r_1)$ is a zero mode,  with $r_1$ determined  by boundary condition at $r=r_0$.}.)
Once  boundary conditions at $r=r_0$ are specified, Eqs. (\ref{eq:scalarform}) for $T$,  $Q_i$ and $S_i$ separately turn into an eigenvalue problem,  with  orthonormal condition:
\be
\int_{r_0}^\infty dr\; r  \;  \Phi_n \; \Phi_m =\delta_{n,m}\;.
\ee
and eigenvalues $m_n^2$ , $n=0,1,\cdots$.  From these, one is led  to a discrete spectrum for $QCD_4$ associated with each bosonic supergravity mode.

\subsection{Parity and Charge Conjugation Assignments}

Next we determine how the supergravity fields and therefore the
glueballs couple to the boundary gauge theory.  This allows us to
unambiguously assign the correct parity and charge quantum numbers to
the glueball states, following the analysis done in \cite{Brower:2000rp}. 

For this purpose, we consider  the
Born-Infeld action plus Wess-Zumino term, describing the coupling of a
supergravity field to  a single D3-brane,
$$S=\int d^4x
\det[G_{\mu\nu}+e^{-\phi/2}(B_{\mu\nu}+F_{\mu\nu})]+\int d^4x (C_0
F\wedge F+ C_2\wedge F+ C_4)\> \; ,$$
where $\mu, \nu\ =  1,2, 3,4$.  We will
find the charge conjugation and parity assignments with the help of
the symmetries of the 4-d gauge theory.  The Euclidean time is taken to be $x_4$ and the spatial
co-ordinates, $x_i$, $i=1,2,3$.

For the 4-d gauge fields, we define parity by
\bea
P&:& A_i(x_i,x_4)\rightarrow  -A_i(-x_i,x_4),\cr
P&:& A_0(x_i,x_4)\rightarrow  A_0(-x_i,x_4).
\eea
for $x_i\rightarrow -x_i$, $x_4\rightarrow x_4$.
 
Charge conjugation for a non-Abelian gluon field is
\be
C: \half T_a A^a_\mu(x) \rightarrow - \half T^*_a A^a_\mu(x)
\ee
where $T^a$ are the Hermitian generators of the group.  In terms of
matrix fields ($A \equiv\half T_a A^a$), $ C: A_\mu(x) \rightarrow
- A^T_\mu(x).$ This leads to a subtlety. For example consider the
transformation of a trilinear gauge invariant operators,
\be C: Tr[ F_{\mu_1 \nu_1} F_{\mu_2 \nu_2} F_{\mu_3 \nu_3}]
\rightarrow - Tr[ F_{\mu_3 \nu_3} F_{\mu_2 \nu_2} F_{\mu_1 \nu_1} ] \; .
\ee
The order of the fields is reversed. Hence the symmetric products,
$d^{abc} F^a_1 F^b_2 F^c_3$, have $C = -1$ and the antisymmetric products,
$f^{abc} F^a_1 F^b_2 F^c_3$,  $C = +1$. Of course using a single
brane, we can only find symmetric products. For reasons explained for instance 
in \cite{Brower:2000rp}, we will only encounter symmetric traces
over polynomials in F, designated by $Sym~\tr[F_{\mu\nu} \cdots]$. Even
polynomials have $C = +1$ and odd polynomials $C = -1$.

\vskip5mm

\noindent{\bf Graviton, Dilaton and Axion States}

Expanding the Born-Infeld action, we
can now read off the $J^{PC}$ assignments. One finds that the graviton $G_{\mu\nu}$ couples as $G_{\mu\nu}T^{\mu\nu}$, where   $T^{\mu\nu}\sim \tr(F_{\mu\lambda}F^\lambda_\nu )
    + \cdots  \; .$
 Because an even number of gluons occur in the field
operators, the charge conjugation for all such states are $C = +$.
For parity, we assume we are in a gauge where the indices of
$G_{\mu\nu}$ do not point along $x_0, r$. From the coupling,
$G_{\mu\nu}\tr[F^{\mu\lambda}F_\lambda^\nu] + \cdots$, we get states
\be
G_{ij}~\rightarrow ~2^{++}\;\;
\ee
The dilaton couples as $\phi \tr{F^2}$, leading
to
\be
\phi~\rightarrow ~0^{++}\;,
\ee
and for the axion coupling $C_0\tr(F_{12}F_{34})$,
\be
C_0 ~\rightarrow ~0^{-+}\;\;  \; .
\ee

\vskip5mm
\noindent{\bf Two-Form Fields}

Consider first the NS-NS 2-form field $B_{\mu\nu}$.  For
$U(1)$ gauge theory in
leading order this field couples as $B_{\mu\nu}F^{\mu\nu}$.
More generally in  the $SU(N)$ gauge
theory,  $tr(F)=0$,  we must have
a multi-gluon coupling, $B_{\mu\nu}\; Sym\tr[ F_{\mu\nu} W ]$, where  $W$ is an
an even power of fields $F$ and the  trace is symmetrized. The first non-trivial coupling for $B_{\mu\nu}$ involves the totally-symmetric    color-singlet  operator,  $Sym\;Tr(F_{\mu\nu}F^2)$,  i.e.,  the d-coupling. 

For parity, again
assume that we are in a gauge where the indices of the 2-form do not
point along $x_4, r$. With $i,j=1,2,3$,
the coupling $B_{ij}\;Sym\tr[F^{ij} W ]\; $
leads to
\be
B_{ij} ~\rightarrow ~1^{+-}\;\; \; ,
\ee

For the Ramond-Ramond 2-form $C_{2,\mu\nu}$,
we have the coupling $ \epsilon^{ijk} C_{2,ij}\; Sym\tr[F_{k 4}W]$, so
\be
C_{2,ij} ~\rightarrow ~1^{--}\; .
\ee
The first non-trivial coupling for  the RR tensor $C_{2,\mu\nu}$  is $Sym\;Tr(\tilde F_{\mu\nu}F^2)$.
\vskip5mm
The complete parity and charge conjugation assignments are now summarized  in
Table~\ref{tab:classIIB} below. 
\vskip5mm
\begin{table}[h]
\begin{center}
\begin{tabular*}{100mm}{@{\extracolsep\fill}||c|c|c|c|c||}
\hline
\hline
$\;\;\; G\;\;\;$ & $\phi$ &$ C_0$   &
$B$    & $C_2$                     \\
\hline
$2^{++}$       &     $0^{++}$    &  $0^{-+}$             &
$1^{+-}$        & $1^{--}$                       \\
\hline
\hline
\end{tabular*}
\caption{$J^{PC}$ Classification for  $QCD_4$ glueball states from IIB Gauge/String Duality. The identification here  differs slightly from that in \cite{Brower:2000rp} where one starts from an 11-d M-theory description with confinement  modeled by an $AdS^7$ black hole metric.}
\label{tab:classIIB}
\end{center}
\end{table}

\subsection{Glueball Spectrum at $\lambda= \infty$}

Let us next  examine the boundary conditions at $r=r_0$. Recall that  the hard-wall model is not a fully consistent theory. However,  it does capture key features
of confining theories with string theoretic dual descriptions.
The main advantage of the hard-wall model is that it can be treated
analytically.
The boundary condition at the wall on the
five-dimensional graviton (and its trajectory for general $J$) is
constrained by energy-momentum conservation in the gauge theory.  We
must impose the boundary condition, $\partial_r(r^{-2}t_{ij})=0$, or, equivalently, Neumann boundary condition for $T(r)$, 
\be
\partial_r T(r) = 0
\ee
at $r=r_0$.  The logic is
the same as in deriving the wave equation $\nabla_P\; h_{ij}=0$: the pure gauge
solution $t(r) = r^2$ must be retained as a zero mode, else conservation of the
energy-momentum tensor will be violated. This condition extends to the
Pomeron for small $|j-2|$, which will be the regime we will mainly consider
below.  
\begin{center}
\begin{table}[th]
\begin{tabular*}{160mm}{@{\extracolsep\fill}||l||c||c|c|c|c||}
\hline
\hline
$J^{PC}$:     & $2^{++} $  &    $0^{++}$ &      $0^{-+}$    &$1^{+-}$  &$1^{--}$
            \\
\hline

Lattice & 1  &0.52 	 & 1.17  & 1.50 	&  2.57   \\
\hline
\hline
Modes:     &$T$     & $S_1$  & $S_2$ &   $Q_1$   &    $Q_2$ \\
\hline
Bdry C.& $\partial_u T |_{0}=0$  &$\partial_u s_1|_{0}=0$ & 
$\partial_u S_2|_{0}=0$  &  $\partial_u q_1|_{0}=0$ &  $\partial_u Q_2|_{0}=0$ \\
\hline
n = 0 &    1.00       &   0.64 &
      1.00  &    1.93 &  2.44 \\
\hline
n = 1 &     3.35      &     3.06&
       3.35 &    5.87&    6.20 \\
\hline
n = 2 &    7.05    &    6.77 &
       7.05 &   10.95 &    11.25 \\
\hline
n = 3 &    12.09    &    11.81 &
     12.09 &    17.36 &    17.65 \\
\hline
\hline
\end{tabular*}
\caption{The mass spectrum, $m_n^2$, $n=0,1,2,3$ for $QCD_4$ Glueballs vs Lattice calculations. }  \label{tab:mass4}
\end{table}
\end{center}
For simplicity, we shall  assume a similar boundary conditions for $Q_i$ and   $S_i$, i.e., with Neumann conditions for all three,
\be
\partial_r T  =\partial_r S_i= \partial_r Q_i=0 \;,  \label{eq:Neumann}
\ee
 at $r=r_0$.   Note, due to degeneracies noted earlier, we have, for the low-mass glueballs,
\be
m^2_0(0^{++})=m^2_0(0^{-+}) = m^2_0(2^{++}) < m^2_0(1^{+-} ) = m^2_0(1^{--})\;.
\ee
 In Table \ref{tab:mass4}, we  have listed mass squared from lattice calculations for ground states, in ratios relative to  $m^2_0(2^{++})\simeq 5.76 {\rm GeV}^2$, e.g. $m^2_0(2^{++})=1$, $m^2_0(0^{++})=.52$, etc. We have  normalized our calculated mass squared with  $m^2_0(2^{++})=1$, and presented  result of this calculation, for $T:\; m^2_n(2^{++})$, $S_2:\; m^2_n(0^{-+})$,  and $Q_2:\;  m^2_n(1^{--})$, $n=0,1,2,3$.    (For two-form fluctuations, we have considered only the case $m^2_{AdS}=16$, since the case of $m^2_{AdS}=0$ can be shown to correspond to pure gauge.) 

In this paper, we are less concerned  on  how well the resulting spectrum agrees with the lattice result and will postpone   further improvements  to future investigations. Nevertheless, it is important to note  the pattern of degeneracy between  the lightest scalar ($0^{++}$) glueball, the lightest tensor ($2^{++}$) and the pseudoscalar ($0^{-+}$), and the degeneracy between two  spin-1 states, ($1^{+-}$ and $1^{--}$), are not shared by  the lattice result.   Within a hard-wall model, these patterns can only be broken by boundary conditions. There is {\it a priori} no reason to adopt the same boundary condition for $S_i$ and $Q_j$, $i,j=1,2$. In fact, based on the results for a black hole background \cite{Brower:1999nj,Brower:2000rp,Constable:1999gb},  it is suggestive that the physics of confinement could be better simulated by having different boundary conditions  for $S_1$ vs. $S_2$, or $Q_1$ vs. $Q_2$, in the IR, thus breaking their degeneracies.

The spectrum can be changed if  different  boundary conditions for $S_1$ and $Q_1$ are adopted~\footnote{ Dirichlet boundary condition was used in \cite{BoschiFilho:2002ta} as a model for calculating the scalar glueball masses. Subsequently, both Dirichlet and Neumann boundary conditions were used  in a holographic approach  to glueballs masses \cite{Boschi-Filho:2005yh}. For related works,  see \cite{brodsky,Karch:2002sh}.}.  We find it amazing that it is possible to adjust the boundary conditions at $r=r_0$, so that $m^2_0\;(0^{++}) < m^2_0\;(2^{++}) \simeq m^2_0 \;(0^{-+})< m^2_0 \;(1^{+-})< m^2_0 \;(1^{--})$. As noted earlier, this pattern is consistent with that followed from a constituent gluon and bag models.
As an illustration, we  adopt Neumann conditions for $h_{ij}$, $s_1$, $S_2$, $q_1$ and $Q_2$. The will change the masses for $0^{++}$ and $1^{+-}$, and the resulting mass squared under Neumann conditions, $S_1:\; m^2_n(0^{++})$,  and $Q_1:\;  m^2_n(1^{+-})$, $n=0,1,2,3$,  are  also shown in Table \ref{tab:mass4}.

\subsection{Regge Trajectories}

At $\lambda$ large but finite, $0(1/\sqrt\lambda)$ corrections must be taken into account.   Recall that, for  the metric fluctuations, the on-shell conditions, $ L_0=\bar L_0=1$,  receive corrections, and, operatorially, can be written as 
\be
\left[ {j-2} + (1/2\sqrt\lambda)\nabla_P\right] t(r) =  \left [ {j-2}  + (1/2\sqrt\lambda) (   -  r\partial_r r\partial_r  +4 - \frac{t}{r^2} )\right] t (r) =0
\ee
For  $j=2$, this leads to Eq. (\ref{eq:h1}), where, by applying  the boundary condition at $r_0$, it turns into  an eigenvalue condition  on $t\rightarrow m_n^2$. For finite $\lambda$, we now have a generalized eigenvalue problem. Since $j$ enters as a parameter, we find the glueball spectrum now can be considered as function of $j$ and $\lambda$, $m^2_n(j,\lambda)$. Equivalently, we can treat $t$ as a parameter, and consider this as an eigenvalue problem in $j$, i.e., we will have discrete spectrum in the $J$-plane. Since each eigenvalue  is  a function of $t$, one arrives at a set  $j_n(t)$, $n=0,1,\cdots$,  each specifying  a Regge trajectory.  Furthermore, this is also a continuous spectrum, corresponding to the BFKL cut. The $J$-plane spectrum is illustrated in Fig. \ref{fig:hardwallSpectrum}, as a graph of $j$ vs. $t$.

It is now straight forward to generalize this to the whole supergravity modes.  Using the variable $u=\log (r/r_0)$, it is possible to express all relevant extensions of Eqs. (\ref{eq:reducedform})  in a standard Schr\" odinger form,
\bea
{\rm Pomeron:}    \quad \quad \quad \quad \quad    \left[   - \partial^2_u +4+ (2\sqrt\lambda) (j-2) -e^{-2u}(t /t_0)\right] t(u)& = & 0  \; , \nn
{\rm Odderon:} \;\; \quad\quad \;\;  \left[  - \partial^2_u   +  m^2_{AdS} + (2\sqrt\lambda) (j-1) -e^{-2u}(t /t_0)\right]q(u) &=&   0\; , \nn
{\rm Scalar:}  \quad\quad\quad  \quad \quad \quad     \left[   - \partial^2_u   +4  + (2\sqrt\lambda) (j-0) -e^{-2u}(t /t_0)\right] s(u)  &=  &    0 \;,  
\eea
with boundary conditions, (\ref{eq:Neumann}), becoming
\be
\partial_u (e^{-2u}  t )  =\partial_u (e^{-2u} s_i) = \partial_u (e^{-2u} q_j)=0 \;,\label{eq:Neumann2}
\ee
 at $u=0$.  We  have also made use of the fact that  $t_0=1/ \alpha_{0}^{\prime}\sqrt{\lambda}    =\Lambda_{QCD}^{2} $ and  $\alpha_{0}^{\prime}\equiv\frac{\alpha'R^{2}}{r_{0}^{2}}.$

The spectra for all three cases are structurally  identical. We can express these equations collectively as
\begin{equation}
\left[-\partial_{u}^{2} + {m^2}_{(\gamma)}(j)  -e^{-2u}(t/t_0)\right]\psi^{(\gamma)}(u)= 0   \label{eq:schro}
\end{equation}
 with 
 \be
{m^2}_{(\gamma)}=2\sqrt{\lambda}(j- j^{(\gamma)})
\ee
where  $j^{(\gamma)}= 2-2/\sqrt\lambda, 1- m^2_{AdS}/2\sqrt\lambda,  - 2/\sqrt\lambda$, for $\gamma$ taking  on $t,q,s$ respectively. Note that $j^{(\gamma)}$ is the location of the BFKL cut for the $t, q,s$ modes respectively.  Note also that ${m^2}_{(t)}(j)$ and ${m^2}_{(q)}(j)$ are the same as ${m^2}_{(+)}(j)$  and ${m^2}_{(-)}(j)$  introduced earlier in Eq. (\ref{eq:DE2}). 
Since one can move from one  equation to another by shift the $J$ value, they lead to same $J$-plane structure except for the intercepts at $t=0$.

Let us examine the form of (\ref{eq:schro}) as a Schr\"odinger equation. At $t\leq0$, the potential is strictly positive and approaching zero, at $u=\infty$. We conclude that   the spectrum consists of 
a continuum and there are no bound states. For $t>0$, on the other hand, there is now an attractive  potential well at $u=0$, and bound states can now be formed, in addition to the continuum at $E_\gamma = 0$. 

Changing back to the variable $z=e^{-u}$, our equation becomes a Bessel equation,
 \be
\left[z^{2}\partial_{z}^{2}+z\partial_{z}+({t}/{t_0})z^{2}-\nu^{2}\right] \psi=0,
\ee
 with 
\be
\nu^{2}=m^2_{(\gamma)}= 2\sqrt{\lambda}( j-j^{(\gamma)} )
\ee
 The solutions
are given by
 \be
\psi(z)=c_1 J_{\nu}(\sqrt{t/t_0}\; z)+ c_{2}Y_{\nu}(\sqrt{t/t_0}\; z),
\ee
where $J_{\nu}$ and $Y_{\nu}$ are the Bessel functions of the first
and second kind respectively. At $z=0,$ Bessel functions of the second
kind are singular, and since we want the solution that is regular
at $z=0$ we can conclude that $c_{2}=0$ and hence
\be
\psi(z)= c_1 J_{\nu}(\sqrt{t/t_0}\; z)
\ee

 We  next impose boundary conditions (\ref{eq:Neumann2}) at $z=1$, leading to an  eigenvalue problem. There are two equivalent ways to proceed. One finds
 \begin{itemize}
 \item{} Given $j>j_{\gamma}$ real, solve spectrum $t_{\gamma,n}(j)$, $n=0,1,\cdots$.
 \item{} Given $t>0$, solve spectrum $j_{\gamma,n}(t)$, $n=0,1,\cdots$.
 \end{itemize}
The $J$-structure is now more robust with the emergence of Regge trajectories at positive $t$. For the leading trajectory in the  $C=+1$ sector, the lowest mass state corresponds to a tensor glueball when the trajectory crosses $j=2$. That is, under  gauge/string duality, the Pomeron can be identified as a ``Reggeized massive graviton''.  There will also be interpolating Regge  trajectories associated with modes identified in Table-\ref{tab:mass4}, e.g., that for $C=-1$.  The generic $J$-plane structure  can be illustrated by that for $C=+1$. (Fig. \ref{fig:hardwallSpectrum}.)  The details on how these trajectories emerge from the  BFKL cut will of course be model-dependent. However, features such as their being asymptotically linear at large positive $t$ are  general.  Since this analysis is nearly identical to that of Ref. \cite{Brower:2006ea} for $C=+1$, we will not repeat it here.

\newpage

\section{Comments}\label{sec:conclusion}

In this paper, we have focused on the $C=-1$ $J$-plane singularities in the large $N_c$ limit from the perspective of gauge/string duality. We find that, while Pomeron emerges as fluctuations of the metric tensor, $G_{MN}$,  Odderons can be associated with that of  anti-symmetric tensor fields, i.e., {{Kalb-Ramond}}  fields \cite{Kalb:1974yc}, in $AdS_5$ background.    We have  demonstrated that the strong coupling conformal Odderons are again
fixed cuts in the $J$-plane, just as the case for $C=+1$. Their intercepts are specified by the
AdS mass squared, $m^2_{AdS}$, for both  Kalb-Ramond fields  $B$ and $C_2$,  (parity degenerate), 
\be
j_0^{(-)}=1- m^2_{AdS}/2\sqrt{\lambda} + O(1/\lambda)\;.
\ee
One solution has $m^2_{AdS, (1)} = 16$, and a second solution has $m^2_{AdS, (2)}=0$.  Thus the situation parallels that found in the weak coupling, as summarized in Table~\ref{tab:intercepts}.  Moreover unlike the case of the Pomeron, for the Odderon both the weak and strong coupling solutions start with $j_0 = 1$ as $\lambda \rightarrow 0 $ and $\lambda \rightarrow \infty$ respectively and decrease away from this limit. 

When confinement deformation is taken into account, the $J$-structure becomes more robust with the emergence of  Regge trajectories  at  positive $t$. Recall that, for the leading trajectory in the $C=+1$ sector, the lowest mass state corresponds to a tensor glueball when the trajectory crosses $j=2$. That is, under gauge/string duality, the Pomeron can be identified as a Reggeized massive graviton.   For $QCD_4$ based on metric which is asymptotically $AdS_5$ in the UV, we have identified the $J^{PC}$ quantum numbers  for  all ground state glueballs,   and they are summarized in Table~\ref{tab:classIIB}. We have also shown that there are  interpolating Regge trajectories associated with these modes.  The generic $J$-plane structure can be illustrated by that for $C=+1$. (Fig. \ref{fig:hardwallSpectrum}.)

Let us next comment on the anticipated glueball states with $J^{PC} = 1^{\pm -}$, lying on the leading $C=-1$ Odderon trajectories.  This is certainly the case for the branch $m^2_{AdS}=16$. However, for the branch $m^2_{AdS}=0$, those two states $J^{PC} = 1^{\pm -}$ decouple from the physical spectrum due to gauge invariance. To be more precise, since the associated field strengths actually vanish, these states do not couple to other physical modes. However, there are at least two good reasons for us to accept these Odderon trajectories associated with the branch $m^2_{AdS}=0$.  First, as one moves away from $j =1$, the system is effectively massive and the field strengths no longer vanish. There will no longer be residual gauge freedom for decoupling and states on the Regge trajectories are now physical. (For this branch, the first physical recurrence along a Regge trajectory occurs at $j=3$.) Second, there is also the possibility of a Higgs-like  mechanism at work so that these spin-1 states will acquire masses non-perturbatively. This has indeed been suggested to be the case when coupling to open strings is taken into account \cite{Kalb:1974yc,Cremmer:1973mg}, generating masses of the order $O(1/g^2)$. From AdS/CFT perspective, however, we are at this point unable to address this possibility. The best we can do is to introduce ``probe branes''. Nevertheless, it does suggest that, when coupling to external hadrons is taken into account, these would be gauge modes could indeed manifest themselves non-perturbatively through a Higgs-like mechanism. However, until this happens, this conformal Odderon, $j^{(-)}_{0,(2)}$, remains at $j=1$. In this connection, it is intriguing to note that, from weak coupling, the Odderon mode with intercept at $j=1$ is also ``anomalous''. Its existence requires ``enlarging'' the Hilbert space of acceptable physical states, e.g., involving delta-function in gluon separations in impact space. These more singular configurations at small impact separations suggest, from string/gauge dual perspective, having more singular behavior at the $AdS$ boundary. This is precisely the case for $m^2_{AdS}=0$, as compared to $m^2_{AdS}=16$.  Possible connections between these observations deserve further examination.  We have also noted that, from strong coupling, $j^{(-)}_{0,(1)}$, corresponds to twist four. This is probably consistent with that from a weak coupling consideration. For $j^{(-)}_{0,(2)}$, on the other hand, no clear conclusion can be drawn at this time, either from weak coupling or from strong coupling. (For weak coupling, see Sec. 3.2.7 of  \cite{Ewerz:2005rg} for a brief discussion.)

Let us re-iterate on the generality of the present Odderon treatment. As is  the case of Pomeron in Gauge/String Duality, the current approach relies on the fact that ${\cal N}=4$ SYM in the strong coupling  can be described by a string theory in $AdS$ background. The dual string theory affords a perturbative treatment at large curvature, and, in this limit, one approaches super gravity in ten dimenison. When viewed as a two-dimensional world-sheet sigma-model, there always exists, in additional to the (symmetric) metric tensor, $G_{MN}$, an anti-symmetric tensor field, $B_{MN}$.  In this paper, we  have shown that, just like the fact fluctuations of $G_{MN}$ can be identified with a dual Pomeron where $C=+1$, fluctuations of $B_{MN}$ leads to dual Odderons, with $C=-1$. Therefore, the generality of the present work is on an equal footing as that for the dual Pomeron~\cite{Brower:2006ea}. Further discussion on the theoeretical foundation for this dual approach can be found in Sec.~\ref{sec:review}.  For historical reasons, these anti-symmetric tensor fields have generically been referred to as ``Kalb-Ramond" fields. As explained in Sec.~ \ref{sec:spectrum}, when  confinement deformation is introduced, small fluctuations of these modes can be identified with glueball states. For the K-R fields, one finds  both $J^{PC}$ equal $1^{+-}$ and $1^{--}$ glueballs.

It is also useful to  provide further contrast between  the current  treatment of odderons with the traditional perturbative approach under a leading log approximation.  We comment on some of issues which normally come up in the later approach and discuss their roles, if any, from gague/string duality. Many more questions can undoubtedly be raised and we limit here to some of the more common ones.

\begin{itemize}
\item
In a perturbative LLA, BFKL Pomeron~\cite{Lipatov:1976zz,Kuraev:1977fs,BL,Lipatov:1985uk,KirschLipat}  can be thought of as a color-singlet bound state of two Reggeized gluons. Gluon Reggeization is an integral step for both the BFKL construct for the Pomeron and the BKP construct~\cite{Kwiecinski:1980wb,Bartels:1980pe} for the Odderon~\footnote{Gluon Reggeization can be characterized by a boostrap condition  from s-channel unitarity perspective, and it can be tested perturbatively. The importance of proving this condition beyond the leading order has been emphasized in a recent publication~\cite{Stasto:2009bc}.}.   In our dual approach, we work with the color-singlets only and the notion of gluon Reggeization does not arise. 

\item In contrast, since the underlying structure in a dual approach  is a string theory, the presence of a $J$-plane is natural. Thus one speaks of a dual Pomeron as a ``Reggeized graviton" in the case of $C=+1$ and Odderon as a ``Reggeized K-R" field for $C=-1$. In the strong coupling, the structue of the $J$-plane can be established semi-classically. As shown in \cite{Brower:2006ea} and in this paper, at $\lambda$ large but finite, the $J$-plane  in the conformal limit  for both $C=\pm 1$ involves branch cuts only.  However, in a confining theory, e.g., for QCD, Regge trajectories will emerge for positive $t$ (Sec.~\ref{sec:spectrum}.)

\item   In our dual approach, energy-momentum conservation  is maintained manifestly. Indeed, the vanishing of the anomalous dimension at $J=2$  plays a central role in establishing the location of the Pomeron intercept in the conformal limit at strong coupling.  (See Sec. \ref{sec:general}.)  Furthermore, Moebius invariance enters as a subgroup of the conformal symmetry and is  now realized as the isometry of the transverse $AdS_3$, as emphasized in \cite{Brower:2007xg,Brower:2007qh} and briefly reviewed in the Appendix \ref{app:conformal}. 
\ignore{ It is interesting to point out that, for the Odderon,  the vanishing of the anomalous dimension at $j=1$ (Lorentz spin $J_L=j+1=2$), follows from charge conservation. In contrast, in extrapolating  a perturbative sum to the strong coupling (where one approaches a spin-2 exchange), it is necessary to modify the traditional BFKL calculation by introducing energy-momentum conservation effects. Under such modifications, it is unclear how some of the previously established properties of the theory such as Moebius invariance are affected. }

\item In earlier  weak coupling BFKL approach, Moebius invariance requires the vanishing of the wave function whenever the separation between any pair of  gluons vanishes. However, for the second Odderon solution in the weak coupling where $j^{(-)}_{0,(2)}=1$, this restriction has to be relaxed by enlarging the Hilbert space of states~\cite{Bartels:2005ji,Stasto:2009bc}.  It is  interesting to note that there is a  hint of a similar feature in a dual appraoch, i.e., the normalizability condition for the $j^{(-)}_{0,(2)}=1$ solution differs  from that of $j^{(-)}_{0,(1)}<1$. Clearly, this subtle point deserves further investigation in a future study.\ignore{  It is also intersting to note that the enlargement of the Hilbert space can manifest by the way how a weak coupling  Odderon couples to external hadrons~\cite{Ewerz:2005rg}. In contrast, in a dual approach, this has to be introduced phenomenologically since, at our current stage of development,  mesons and baryons are treated separately from glueballs in AdS/CFT.  }

\item In a weak coupling  treatment, Pomeron is minimally a bound state of 2 Reggeized gluons and Odderon a bound state of three.  Indeed, in a weak coupling, there are  additional Pomerons and Odderons involving more than the minimal numbers of gluons. For each sector with a fixed number of gluons, the solution can in principle be found by exploiting the integrabie structure of a Heisenberg spin chain of a finite length. In our dual approach, both the Pomeron and the Odderon are collective modes of gluons in color singlet configurations. In this limit, they each can be represented by fluctuations of local fields, $G_{MN}$ and $B_{MN}$ respectively.  The notion of a fixed number of constituent gluons losses its meaning. \ignore{It is interesting to speculate that the strong coupling results corresponds to some ``mean field" solution of  ``long" Heisenberg spin chains.}

\item  Lastly, let us comment on the representation labels for the SL(2,C). One of more interesting features of the weak coupling approach is the holomorphic and anti-holomorphic separability and their correspondence to the unitary principal series representation of $SL(2,C)$,  labelled by a real and an integer $n$. For the usual leading weak coupling BFKL Pomeron, one has $n=0$, although other $|n|=1,2, \cdots$ solutions are also allowed. When $n\neq 0$, angular correlation between color dipoles enters. As stressed in \cite{Brower:2007xg}, our strong coupling solution corresponds to a realization where $SL(2,C)$ enters as the isometry of transverse $AdS_3$ and only $n=0$ solution appears.  This is likely a  reflection of the fact that, in strong coupling,  effective local fields can be used.  While one is able in strong coupling to encode ``parton size", orientational information such as that of  ``color dipole" no longer enters in the formulation~\cite{Brower:2006ea,Brower:2007xg}. How such orientational information can be encoded  in the strong coupling limit is an intersting question,  but it is beyond the scope of our current analysis.  
  \end{itemize}

 Let us next turn to a brief comment on eikonalization.  Consider elastic scattering $a+b \to a+b$ and $\bar a + b \to \bar a +b$. If the
 $C=-1$ contribution can be neglected,  it follows from  Refs.\cite{Brower:2007xg,Brower:2007qh,Cor} that,
  in the strong coupling limit,  $F(s,t) \simeq \bar F(s,t)\simeq  F^{(+)}(s,t)$ can be expressed in
  a ``generalized'' eikonal representation over  $AdS_3$, 
\be
F(s,t) =\int dz dz' P_{13}(z) P_{24}(z') \int d^2b \; e^{-ib^\perp q_\perp}  \widetilde A(s,b^\perp,z,z')\;, \label{eq:eikonalization}
\ee
where
\be
\widetilde A(s,b^\perp,z,z')= 2 i s \left [ 1-e^{i\chi(s,b^\perp, z,z')} 
\right] \;.
\ee
and $b^\perp = x^\perp - x'^\perp$ due to
translational invariance.   The probability distributions for left-moving, $P_{13}(z)$, and right moving, $P_{14}(z)$ particles are products of initial (in) and final (out)
particle wavefunctions:
\be
P_{13}(z) =( z/R)^2\sqrt{g(z)}  \Phi_1(z) \Phi_3(z) \quad \mbox{and} \quad
P_{24}(z) = (z'/R)^2\sqrt{g(z')}  \Phi_2(z') \Phi_4(z') \; .
\ee
When confinement is implemented, wave functions can be normalized so that $\int dz P_{ij}(z)=\delta_{ij}$. 
By expanding to first order in the coupling $g_0^2$, this eikonal can then be related to the transverse representation for the strong coupling Pomeron kernel,
\be
\chi(s,x^\perp- x'^\perp,z,z')=
  \frac{ g_0^2 R^4 }{ 2(zz')^2 s}  {\cal K}^{(+)}(s,x^\perp - x'^\perp,z,z') \; ,
\ee
with  ${\cal K}^{(+)}$  given by (\ref{eq:PomeronOdderonkernel}). This kernel was first introduced in Ref. \cite{Brower:2006ea}, where the dimensionless
coupling $g_0^2$ is proportional to the $AdS_5$ gravitational coupling
constant: $g_0^2\sim \kappa_5^2/R^3\sim 1/N^2$.  This is a natural
generalization of our earlier result for $AdS$ graviton
exchange~\cite{Brower:2007qh,Cor}, whose kernel can be obtained by
taking the limit $\lambda \rightarrow \infty$.  

 When both $C=\pm 1$ are present,  eikonalization can still be carried out by generalizing the analysis of \cite{Finkelstein:1989mf}, e.g., the amplitude $F$ still takes on the form, Eq. (\ref{eq:eikonalization}), with a new  eikonal $\chi = \chi^{(+)} + \chi^{(-)}$, and the same for crossed amplitude, $\bar F$, with eikonal $\bar \chi =  \chi^{(+)} - \chi^{(-)}$.  The $C=-1$ eikonal  is now given by the corresponding Odderon kernel,
\be
\chi^{(-)}(s,x^\perp- x'^\perp,z,z')=
-  \frac{ e_0^{(a)}e_0^{(b)} R^4 }{ 2(zz')^2 s}  {\cal K}^{(-)}(s,x^\perp - x'^\perp,z,z') \; .
\ee
with ${e_0}^{(a)}$ an effective coupling constant.    It then follows from (\ref{eq:eikonalization}) that
\be
\; F^{(\pm)}(s,t) =\int dz dz' P_{13}(z) P_{24}(z') \int d^2b \; e^{-ib^\perp q_\perp}  \widetilde A^{(\pm)}(s,b^\perp,z,z')\;,
\label{eq:eikonal_odderon}
\ee
where
\bea
\widetilde A^{(+)}(s,b^\perp,z,z') & =&  2 i s\;  \left [ 1- e^{i\chi^{(+)}(s,b^\perp, z,z')} \cos   \chi^{(-)}(s,b^\perp, z,z')\right]\;,  \label{eq:evenpartialwave} \\
\widetilde A^{(-)}(s,b^\perp,z,z') & =&  - 2  s  \; e^{i\chi^{(+)}(s,b^\perp, z,z')} \sin   \chi^{(-)}(s,b^\perp, z,z') \;.\label{eq:oddpartialwave}
\eea

Let us briefly  comment on the issue of saturation and Froissart bound.  Saturation for $F$ and $\bar F$ is   characterized by the condition $\chi \simeq O(1)$ and $\bar \chi = O(1)$ respectively. With $j^{(+)}_0>j_0^{(-)}$, the corrections from the  $C=-1$  contribution is a higher order effect.  As also pointed out in \cite {Brower:2007xg}, Froissart-like behavior for the $C=+1$ sector requires confinement, which leads to a diffractive disk, with a radius $r_0(s) \sim \log s$. This remains the case when $C=-1$ is present. This can be explicitly verified by examining Eq. (\ref{eq:evenpartialwave}).  

The analog of the Froissart bound for
the $C=-1$ sector  is $\Delta \sigma_T(s) \sim \log s$, which if saturated
\cite{Lukaszuk:1973nt,Bialkowski:1974cp,Finkelstein:1989mf,Avila:2006wy,Nicolescu:2007ji},   has been referred to as the ``Maximal Odderon''.  A natural question one could raise is whether such a behavior would emerge under a similar eikonal consideration.   With $j^{(-)}_0<1$, it follows from Eq. (\ref{eq:oddpartialwave}) that $\Delta \sigma$ vanishes at least as fast as $s^{j_0^{(-)}-1}$, and  the bound on $\Delta \sigma$ is far from being saturated. Moreover, even if we assume that $j^{(-)}_0>1$, unlikely as it maybe, the conclusion is again no ``Maximal Odderon'' saturation is
found.

In flat space, it has been shown~\cite{Finkelstein:1989mf} earlier  that eikonal sum does not lead to the Maximal Odderon saturation. The flat space argument follows from  (\ref{eq:oddpartialwave}) with
the z-coordinate fixed in the IR.  Because of confinement there is a diffractive disk of radius $r_0(s)\sim \log s$. The corresponding flat-space eikonal factor $e^{i\chi^{(+)}(s,b^\perp)}$ is either highly oscillatory or absorptive within this disk. In either case, it leads to rapid suppression,  yielding negligible contribution to $\Delta \sigma$. With $j^{(-)}_0>0$, scattering for $F^{(-)}$ can take place within  a ``ring'' at the edge of the diffractive disk,  of width $\delta r$, leading to a finite contribution to $\Delta \sigma$, of the order $ r_0(s) \; \delta r \sim  \log s\;  \delta r$.  However, one can also show that the width of this ring decreases as $s^{-\Delta j}$, where $\Delta j = j^{(+)}_0-j^{(-)}>0$. Therefore, upto a logarithm, $\Delta\sigma$ vanishes as a power of $s$, within the eikonal picture. Essentially the same argument goes through for the eikonal
expansion (\ref{eq:oddpartialwave}) of the AdS hardwall model  because
the z-coordinate is constrained to a  finite interval.

As summarized in \cite{Ewerz:2005rg}, the strongest evidence for Odderon, at this point, is more theoretical than experimental. Our current analysis has provided further support for its existence from a non-perturbative perspective. Nevertheless, puzzles remain on why it is so difficult to observe  the  Odderon experimentally. 
It has been suggested that direct exclusive production holds the best chance for observing Odderon. One would expect that at $t =0$
the Odderon would dominate over the quark anti-quark (meson) Regge trajectories and that for large $t <0$, there would
be a distinguishing hard components similar to that
expected for the BFKL Pomeron.  If so, better understanding and control of
the Odderon intercept and its on-shell string vertex-coupling would be useful. Our strong coupling approach might be able to shed more light on this question. This and other related issues will be addressed in a subsequent investigation.

\newpage
{\bf Acknowledgments}:  We would  like to thank S. Mathur, J. Polchinski and M. Strassler for earlier collaborations which are instrumental for the completion of this work. Thanks should also go to  E. Levin,
Y. Kovchegov, C. Marquest, C. Ewerz,, L. Lipatov, I. I. Balitsky, B. Nicolescu, P. Gauron, A. B. Kaidalov, H. M. Fried, L. R. A. Janik, J. Wosiek,  J. Bartels, M. Lublinsky, A. Kovner, L. McLerran and many others for discussions on various aspects of Odderon physics.   We would also like to remember  K. Kang and J. Kwiecinski for their influence on our views  of the Odderon as well as on other aspects of high energy hadronic collisions. 
RCB and CIT would like to  thank the Galileo Galilei Institute for Theoretical Physics for the hospitality and the INFN 
for partial support during the final stage of completion of  this work. RCB's research  is supported in part by the Department of 
Energy under
Contract.~No.~DE-FG02-91ER40676. 
MD's and   CIT's research are
supported in part by 
the U.~S.~Department of 
Energy under  Grant DE-FG02-91ER40688, TASK A.

\newpage

\def\theequation{A.\arabic{equation}}
\setcounter{equation}{0}

\appendix

\newpage
\section{Wave Equations}

We outline the derivation of the wave equations that
were used to find the energy levels in the supergravity theory.   The discussion here is similar to that in \cite{Brower:2000rp}, except we now work with the $AdS_5$ metric, (\ref{AdSWmetric}), with $R=1$ for simplicity,  and a hard-wall cutoff.

The simplest equation is the scalar wave equation for the dilaton and the
axion. For the transverse traceless metric fluctuations, one can simply work with the linearized Einstein equation. Both of these can be found in \cite{Brower:2000rp}. Let us focus here on the case of the two-form fields,
$B_{\mu\nu}$ and $C_{\mu\nu}$, which at
the linear level are conveniently combined into
a single complex field, $\tilde B_{\mu\nu} =B_{\mu\nu} + i
C_{\mu\nu}$.  (We drop the subscript $2$ of $C_{2,\mu\nu}$ in what follows to simplify the writing.) Let us introduce  a Maxwell operator for two-forms
\bea
(\square_{Maxwell} \; \tilde B)_{\alpha\beta}&=&\frac{3}{\sqrt{-g}}g_{\alpha\mu}g_{\beta\nu}\partial_{\lambda}(\sqrt{-g}g^{\lambda\lambda'}g^{\mu\mu'}g^{\nu\nu'}\partial_{[\lambda'}\tilde B_{\mu'\nu']}) \; , 
\eea
or, equivalently,
\bea
(\square_{Maxwell} \; \tilde B)_{\alpha\beta}
&=&  g_{\alpha\mu}g_{\beta\nu}(g^{\mu\mu'}g^{\nu\nu'}\tilde H_{{\mu'}{\nu'}\lambda}){}^{;\lambda}\;, 
\label{eq:Maxwell}
\eea
where 
\be
\tilde H_{\mu\nu\lambda}=\partial_{\mu}\tilde B_{\nu\lambda}+\partial_\nu
\tilde B_{\lambda\mu}
+\partial_\lambda \tilde B_{\mu\nu}\; ,
\ee
is the associated  field strength. 
For ease of writing, we adopt the standard shorthand: $\square_{Maxwell} \; \tilde B_{\alpha\beta}$ for $(\square_{Maxwell} \; \tilde B)_{\alpha\beta}$.

 It was shown in
\cite{Kim:1985ez} that this field satisfies the equation
\be
(\square_{Maxwell}-k(k+4))\tilde
B_{\mu\nu}+2i\epsilon_{\mu\nu}{}^{\rho\lambda\sigma}\partial_\rho
\tilde B_{\lambda\sigma}=0\; ,
\ee
for  $k=0,1,\cdots$ harmonics on $S^5$.
Since the second order differential operator
factorizes into two first order operators,
solutions fall into two classes,
\bea
i(*D \tilde B^{(1)})_{\mu\nu} &=& 2(k+4)\tilde B_{\mu\nu}^{(1)} ,\nn
  - i (*D \tilde B^{(2)})_{\mu\nu} &=& 2k \tilde B_{\mu\nu}^{(2)}  \; ,\label{eq:linearB}
\eea
where $(*D \tilde B)_{\mu\nu}=\epsilon_{\mu\nu}{}^{\rho\lambda\sigma}\partial_\rho
\tilde B_{\lambda\sigma}$ and $I$ is the identity matrix. It is convenient to iterate these first order equations to get the second order equations,
\bea
(\square_{Maxwell} -(k +4)^2)\tilde B_{\mu\nu}^{(1)}&=&0  \;, \nn
(\square_{Maxwell} -  k^2)\tilde B_{\mu\nu}^{(2)}&=&0\; . \label{eq:quadraticB}
\eea

\subsection{Counting Modes}

To count the number of independent fluctuations for a supergravity field, we imagine harmonic plane waves propagating in the $AdS$ radial direction, $r$, with Euclidean time, $x_4$. For example, the metric fluctuations in $AdS_5$
\be
G_{\mu\nu}= \bar G_{\mu\nu} + h_{\mu\nu}(x)
\ee
in the fixed background $\bar G_{\mu\nu}$ are taken to be of the form $h_{\mu\nu}(r,x_4)$. There is no dependence on the other spatial coordinates, $\vec x= (x_1,x_2,x_3)$. 

\noindent{\bf Metric Fluctuations}

A graviton has two polarization indices. If we were in flat space-time, we could go to a gauge where these indices took on values among $(x_1,x_2,x_3)$ and not from the set $(x_4, r)$. The polarization tensor should also be traceless. This leaves $(3\times 4)/2 - 1=5$ independent components. In the $AdS$ space-time, we can count the number of graviton modes the same way, though the actual modes that we construct will have this form of polarization only at $r\rightarrow \infty$; for finite $r$, other components of polarization will be constrained to acquire nonzero values \cite{Brower:1999nj,Brower:2000rp}.  Therefore, a set of {\it five independent } polarizations can be characterized by the following non-vanishing components at $r\rightarrow \infty$, 
\be
h_{ij}: \quad h_{ij}=h_{ji}\;, \quad {\rm and} \quad Tr\;h=0\;, \quad i=1,2,3\;.
\ee
 These five states form a spin-2 representation under $SO(3)$ as indicated in Table \ref{tab:classIIB}.  Let us consider perturbations of the following form:
\be
h_{ij}(r,x_4) = h_{ij} \; t(r) \; e^{-m x_4}   
\ee
where $i,j=1,2,3$, with  $h_{ij}$ an arbitrary constant traceless-symmetric $3\times 3$ tensor. 
 From linearized Einstein's equations about the $AdS_5$ background, one finds
 \be
[ - {r} \frac{d}{dr} r \frac{d}{dr}   + 4 ] t(r) = \frac{m^2}{r^2} t(r) \;.  
\ee
leading to  Eq. (\ref{eq:h1}).

\noindent{\bf Scalar Fields}

There are fluctuations for the dilaton $\phi$ and the axion $C_0$.  These can  be combined to form a complex field $\tilde \Phi = e^{-\bar \phi - \phi} + i C_0\simeq e^{-\bar \phi}(1-\phi )+ i C_0$. There is also a scalar $G_\alpha^\alpha$ associated with the volume fluctuations in $S^5$, (with $m_{AdS}^2=32$). However, we will not consider this mode here.   

Let us consider perturbations of the following form:
\bea
\phi(r,x_4) &=& S_1(r) \;e^{ik_4x_4}     \nn 
C(r,x_4) &=& S_2(r) \; e^{ik_4x_4}  \; .         
\eea
and re-write $k_4=im$. From the scalar Laplacian, one finds, for $i=1,2$, 
 \be
    -  \frac{d}{dr} r^5 \frac{d}{dr} S_i(r)= {m^2}\; {r}\; S_i(r),  
  \ee
leading to Eq. (\ref{eq:scalarform}).

\noindent{\bf Kalb-Ramond Two-Form  Fields}

 As explained earlier, the field equation for $\tilde B$ can be factorized into two first order equations, and each can be iterated leading to a second order equation of the form
\be
\square_{Maxwell}\; \tilde B_{\mu\nu} + m^2_{AdS,i} \tilde B_{\mu\nu}=0\;.
\ee
where $m^2_{AdS,1}= (k+4)^2$ and $m^2_{AdS,2}=k^2$, $k=0,1,\cdots.$  For the purpose of counting modes, polarizations for massless 2-form in $AdS_5$ can also be restricted to be transverse, with fields depending only on $(x_4,r)$. Therefore, the polarization tensor is an antisymmetric 2-tensor in the direction $x_1,x_2,x_3$, leading to 3 independent components. On the other hand, since we in general need to work with massive 2-form, longitudinal modes are allowed and it would appear that the polarization tensor is an antisymmetric tensor in coordinates $x_1, x_2, x_3,r$, with 6 independent components. 
This assertion turns out not to be the case, and, the number of independent (complex) components for our massive 2-form, $\tilde B_{\mu\nu}$ is also 3, as if we are dealing with a massless case. This less than intuitive fact follows from the first order relation,  relating real and imaginary parts and leading to further constraints. 

One must  check that solutions to
the second order equation for $\tilde B_{\mu\nu}$, actually are
valid solution to the original wave equation.  For
example, consider the ansatz
\be
B_{12}=-B_{21}=b_{12} Q_1(r)r^2 e^{ik_4x_4}\quad {\rm and} \quad C_{12}=-C_{21}=0\;.
\ee
Once we have a solution for
$ B_{12}$, the first order equations will determine $C_{34}=-C_{43}\neq 0$ and 
$C_{r 3}=-C_{3r}\neq 0$, while allowing all other components of $B$ and $C$ to be zero.  This does not place any constraints on
$ B_{12}$ itself. From Eq. (\ref{eq:quadraticB}),  $Q_{1}$ now  satisfies Eq. (\ref{eq:scalarform}).  A similar ansatz 
\be
C_{12}=-C_{21}=c_{12} Q_2(r)r^2 e^{ik_4x_4}\quad {\rm and} \quad B_{12}=-B_{21}=0\;.
\ee
would lead to  Eq. (\ref{eq:scalarform}) for $Q_2$, and  $B_{34}=-B_{43}\neq 0$ and 
$B_{r 3}=-B_{3r}\neq 0$, while allowing all other components of $B$ and $C$ to be zero.   

From these examples, one finds that  the number of
independent complex  tensor fields, $\tilde B_{\mu\nu}$, is reduced from 6 to 3.  Independent components can be chosen at $r\rightarrow \infty$,
\bea
B_{ij}&: & \quad B_{ij}=- B_{ji}\;, \quad i,j=1,2,3, \nn
 C_{ij}&:& \quad C_{ij}=- C_{ji}\;, \quad i,j=1,2,3,
\eea
each corresponds to spin-1 under $SO(3)$, as  indicated in Table \ref{tab:classIIB}.

There still remain two cases to consider, (1) $m^2_{AdS}=(4+k)^2$, and $m^2_{AdS}=k^2$.  Since we do not consider fluctuations in $S^5$, we can restrict to $k=0$, i.e., we can restrict to $m^2_{AdS,1}=16$ and $m^2_{AdS,2}=0$. As pointed in \cite{Brower:2000rp}, the case of $m^2_{AdS,2}=0$  can be gauged away. Therefore, the state with $j=1$ associated with this mode decouples. However, at finite but large $\lambda$, Regge recurrences will be developed. As one moves away from $j=1$, Regge trajectory associated with this mode should survive~\footnote{A related issue in flat-space was first discussed in the original work by Kalb and Ramond, \cite{Kalb:1974yc} and also  in  \cite{Cremmer:1973mg}. By invoking coupling to open strings, this would be spin-1 state acquires a mass through a Higgs-like mechanism.}.

 \subsection{Low Dimensional Gauge Invariant Operators}
 \label{sec:constituent}

We note that the basic idea behind the AdS/CFT correspondence in the context of
glueballs is similar to an observation made much earlier by Fritzsch
and Minkowski~\cite{Fritzsch:1975tx}, by Bjorken~\cite{Bjorken:1979hv} and by Jaffe, Johnson
and Ryzak~\cite{Jaffe:1985qp}. Namely that the low mass glueball spectrum
can be qualitatively understood in terms of local gluon interpolating
operators of minimal dimension.  For example, Ref.~\cite{Jaffe:1985qp} lists all gauge invariant operators for
dimension $\Delta = 4, 5$ and $6$. Eliminating operators
that are zero by the classical equation of motion and states that
decouple because of the conservation of the energy momentum tensor,
the operators are in rough correspondence with all the low mass
glueballs states, as computed in a constituent gluon~\cite{Kaidalov:1999de,Kaidalov:2005kz} or bag
model. Indeed more recently Kuti~\cite{Morningstar:1998da} has pointed out that a
more careful use of the spherical cavity approximation even gives a
rather good quantitative match to the lowest 11 states in the lattice
spectrum.   Consequently it is interesting to compare this set of operators
with the supergravity model. We list below all the operators for $\Delta
\le 6$, based on our cutoff $AdS_5$,  except the operators with explicit derivatives (e.g.  $Tr[F D
F]$) and $Tr[F DD F]$ ):

\begin{center}
\begin{tabular}{|l|r|l|r|}
\hline
Dimension     & State  $J^{PC}$      & Operator            &  Supergravity \\
\hline
$\Delta = 4 $     & $0^{++}$             & $Tr(FF) = \vec E^a\cdot \vec E^a - \vec
B^a\cdot \vec B^a$   & $\phi$ \\
$\Delta = 4 $   &  $2^{++}$               &$T_{ij} =   E^a_i\cdot  E^a_j + B^a_i
\cdot   B^a_j - \mbox{trace}$  & $G_{ij}$ \\
$\Delta = 4 $   &  $0^{-+}$             & $Tr(F\tilde F) = \vec E^a\cdot \vec B^a $ &
$C_0$ \\
$\Delta = 6 $    &  $1^{+-}$     & $Tr(F_{\mu\nu}
\{F_{\rho\sigma},F_{\lambda\eta}\})
\sim d^{abc}F^a F^b F^c $ & $B_{ij}$ \\
$\Delta = 6 $    &  $1^{--}$     & $Tr(\tilde F_{\mu\nu}
\{F_{\rho\sigma},F_{\lambda\eta}\})
\sim d^{abc}\tilde F^a F^b F^c $ & $C_{2,ij}$ \\
\hline
\end{tabular}
\end{center}
In this table we have used a Minkowski metric.  The column
on the right lists the supergravity mode that couples  to each operator.   It is comforting to point out that our solutions for Kalb-Ramond modes, $B$ and $C_2$, at $m^2_{AdS}=16$, precisely lead to $\Delta=6$ at $j=1$.

\newpage

\def\theequation{B.\arabic{equation}}
\setcounter{equation}{0}
\section{Conformal Geometry at High Energies}
\label{app:conformal}

In the context of the AdS/CFT correspondence, it is
important to consider the boost operator relative to the full
$O(4,2)$ conformal group, which are represented as isometries of
$AdS_5$.  The conformal group O(4,2) has 15 generators:
\be
P_\mu,M_{\mu\nu}, D, K_\mu\; .
\ee
Of these, $M_{+-}$ is the generator for the longitudinal boost. In terms of transformations on
light-cone variables, there are two interesting 6 parameter
subgroups: The first is the well known collinear group $SL_L(2,R)
\times SL_R(2,R)$ used in DGLAP. The second is $SL(2,C)$ (or M\"obius invariance used in solving the
weak coupling BFKL equations)  with six generators
\be
 i D \pm M_{12} \; ,\;  
P_1 \pm  i P_2 \; , \;  K_1 \mp i K_2 \;,
\ee
corresponding to the isometries of the Euclidean (transverse) $AdS_3$
subspace of $AdS_5$.
Indeed $SL(2,C)$ is the subgroup generated by all elements of the
conformal group that commute with the boost operator, $M_{+-}$ and as
such plays the same role as the little group which commutes with the
energy operator $P_0$.

Eq.  (\ref{eq:DE2})  is an ordinary differential equations, diagonal in $t$.   As explained  in Ref. \cite{Brower:2007xg}, it is useful to transform the equation  to an impact representation  via two-dim Fourier transform,   $\int dq_\perp e^{-i q_\perp \cdot  x^\perp}$, where $t=-q^2_\perp$.   In fact, it is more convenient to scale $G^{(\pm)}$ by one   factor of $zz'$. One finds, after taking Fourier transform, 
\be
(zz') G^{(\pm)} (j) \rightarrow 
 G^{(\pm)}_3(j,x_\perp-x'_\perp, z,z') \; .
\ee
where
\be
\left [  -z^3\dd_z z^{-1} \dd_z   - z^2  \nabla^2_{x_\perp}  + m^2_{\pm}(j)-1  \right] 
G^{(\pm )}_3(j,x_\perp-x'_\perp, z,z' )  =  z^3\delta(z-z')\delta^2(x_\perp-x'_\perp)  \label{eq:G3}
\ee
This is now in the form of a DE for  $AdS_3$ scalar propagator.  Therefore $G^{(\pm)}_3$ can be thought of as $AdS_3$ Green's functions with $AdS_3$ mass squared, $m^2_{\pm}(j)-1$,  and they  have  simple closed-form
expressions   as a function of the $AdS_3$ chordal
distance  \cite{Brower:2007xg}.

In order to gain a better understanding on the emergence of the $AdS_3$
propagators, let us begin with the Euclidean $AdS_3$ metric, 
\be
ds^2 = \frac{R^2}{z^2} [ dz^2 + dx_1 dx_1 + dx_2 dx_2  ] = ds^2 = 
\frac{R^2}{z^2} [ dz^2 + dw d\bar w] \; ,
\ee
where the transverse subspace is three dimensional, $(w = x_1 +i x_2,\;z)$. The
generators of the $SL(2,C)$ isometries of $AdS_3$ are
\bea
J_0 &=& w \dd_w  + \Half  z\dd_z  \quad, \quad J_- = - \dd_w  \quad, 
\quad J_+ = w^2 \dd_w  + w z\dd_z - z^2 \dd_{\bar w} \nn
\bar J_0 &=& {\bar w} \dd_{\bar w}  + \Half z\dd_{z}  \quad, \quad \bar  
J_- = - \dd_{\bar w}  \quad, \quad  \bar J_+ = {\bar w}^2 \dd_{\bar w}  
+  {\bar w}  z\dd_{z} -  z^2\dd_w  \label{eq:sl2c} \; .
\eea
In the conformal group, this corresponds to the identification,
\bea
 J_0\; , \; J_+\; , \; J_-  &\leftrightarrow&  ( - i D + M_{12})/2\; , \;(P_1 + i P_2)/2\; , \;(K_1 -iK_2)/2 \nn
 \bar J_0\; , \; \bar J_+\; , \; \bar J_-  &\leftrightarrow&  (- i D - M_{12} )/2\; , \;(P_1 - i P_2)/2\; , \;(K_1 + iK_2)/2 \nonumber \; ,
\eea
so that the non-zero commutators in $SL(2,C)$  must be
$[J_0,J_\pm] = \pm J_\pm$, $[J_+, J_-] = 2 J_0$, and $[\bar J_0, \bar J_\pm]
= \pm \bar J_\pm$, $[\bar J_+, \bar J_-] = 2 \bar J_0$.  In general,
unitary representations of $SL(2,C)$ are labeled by $h = i
\nu + (1+n)/2 $, and $\bar h = i \nu +(1-n)/2$, which are the eigenvalues 
for the highest-weight state of $J_0$ and $\bar
J_0$.  The principal series is given by real $\nu$ and integer $n$.
The quadratic Casimirs $J^2$ and $\bar J^2$ 
have eigenvalues $h(h-1)$
and $\bar h (\bar h -1)$ respectively. In the representation
(\ref{eq:sl2c}) they are 
\be
J^2 = J^2_0 + \Half (J_+J_- + J_- J_+) = \frac{1}{4}[ z^3\dd_z z^{-1} \dd_z   + 4z^2 \dd_w\dd_{\bar w}]  \label{eq:J2}
\ee
with $\bar J^2 = \bar J^2_0 + \Half (\bar J_+\bar J_+ + \bar J_- \bar
J_+) = J^2 $ in this representation.

\newpage
\bibliographystyle{utphys}
\bibliography{odderonDec1}

\end{document}